# Reachability of Communicating Timed Processes*

L. Clemente[1], F. Herbreteau[1], A. Stainer[2], and G. Sutre[1]

[1] Univ. Bordeaux, CNRS, LaBRI, UMR 5800, Talence, France.
[2] University of Rennes 1, Rennes, France.

**Abstract.** We study the reachability problem for communicating timed processes, both in discrete and dense time. Our model comprises automata with local timing constraints communicating over unbounded FIFO channels. Each automaton can only access its set of local clocks; all clocks evolve at the same rate. Our main contribution is a complete characterization of decidable and undecidable communication topologies, for both discrete and dense time. We also obtain complexity results, by showing that communicating timed processes are at least as hard as Petri nets; in the discrete time, we also show equivalence with Petri nets. Our results follow from mutual topology-preserving reductions between timed automata and (untimed) counter automata.

## 1 Introduction

*Communicating automata* are a fundamental model for studying concurrent processes exchanging messages over unbounded channels [21,11]. However, the model is Turing-powerful, and even basic verification questions, like reachability, are undecidable. To obtain decidability, various restrictions have been considered, including making channels unreliable [3,13] or restricting to half-duplex communication [12] (later generalized to mutex [16]). Decidability can also be obtained when restricting to executions satisfying additional restrictions, such as bounded context-switching [19], or bounded channels. Finally, and this is the restriction that we consider here, decidability is obtained by constraining the communication topology. For communicating finite-state machines (CFSMs), it is well-known that reachability is decidable if, and only if, the topology is a polyforest [21,19]; in this case, considering channels of size one suffices for deciding reachability.

On a parallel line of research, *timed automata* [8] have been extensively studied as a finite-state model of timed behaviours. Recently, there have been several works bringing time into infinite-state models, including *timed Petri nets* [9,4], *timed pushdown automata* [2], and *timed lossy channel systems* [1]. In this paper, we study *communicating timed processes* [18], where a finite number of timed automata synchronize over the elapsing of time and communicate by exchanging messages over unbounded channels. Note that, when processes can synchronize,

* This work was partially supported by the ANR project Vacsim (ANR-11-INSE-004).

runs cannot be re-ordered to have uniformly bounded channels (contrary to polyforest CFSMs). For example, consider two communicating processes $p$ and $q$, where $p$ can send to $q$ unboundedly many messages in the first time unit, and $q$ can receive messages only after the first time unit has elapsed. Clearly, all transmissions of $p$ have to occur before any reception by $q$, which excludes the possibility of re-ordering the run into another one with bounded channels.

We significantly extend the results of [18], by giving a complete characterization of the decidability border of reachability properties w.r.t. the communication topology. Quite surprisingly, we show that despite synchronization increases the expressive power of CFSMs, the undecidability results of [18] are not due to just synchronous time, but to an additional synchronization facility called *urgency* (cf. below). Our study comprises both dense and discrete time.

*Dense time: Communicating timed automata.* Our main result is a complete characterization of the decidability frontier for communicating timed automata: We show that reachability is decidable if, and only if, the communication topology is a polyforest. Thus, adding time does not change the decidability frontier w.r.t. CFSMs. However, the complexity worsens: From our results it follows that communicating timed automata are at least as hard as Petri nets.[3]

Our decidability results generalize those of [18] over the standard semantics for communicating automata. In the same work, also undecidability results are presented. However, they rely on an alternative *urgent semantics*, where, if a message can be received, then all internal actions are disabled: This provides an extra means of synchronization, which makes already the very simple topology $p \to q \to r$ undecidable [18]. We show that, without urgency, this topology remains decidable.

Here, we do not consider urgency directly, but we rather model it by introducing an additional *emptiness test* operation on channels on the side of the receiver. This allows us to discuss topologies where emptiness tests (i.e., urgency) are restricted to certain components. We show that, with emptiness tests, not only the topology $p \to q \to r$ is undecidable, as in [18], but also $p \to q \leftarrow r$ and $p \leftarrow q \to r$. Thus, we complete the undecidability picture for dense time.

All our results for dense time follow from a mutual, topology-preserving reduction to a discrete-time model (discussed below). Over polyforest topologies, we reduce from dense to discrete time when no channel can be tested for emptiness. Over arbitrary topologies, we reduce from discrete to dense time, even in the presence of emptiness tests. While the latter is immediate, the former is obtained via a *Rescheduling Lemma* for dense-time timed automata which is interesting on its own, allowing us to schedule processes in fixed time-slots where senders are always executed before receivers.

*Discrete time: Communicating tick automata.* We provide a detailed analysis of communication in the discrete-time model, where actions can only happen at integer time points. As a model of discrete time, we consider *communicating tick automata*, where the flow of time is represented by an explicit tick action: A process evolves from one time unit to the next by performing a tick action, forcing all the other processes to perform a tick as well; all the other actions are asynchronous. This model of discrete-time is called *tick automata* in [15], which is related to the *fictitious-time* model of [8].

[3] And probably exponentially worse, due to a blow-up when translating from dense to discrete time.

We provide a complete characterization of decidable and undecidable topologies for communicating tick automata: We show that reachability is decidable if, and only if, the topology is a polyforest (like for CFSMs), and, additionally, each weakly-connected component can test at most one channel for emptiness. Our results follow from topology-preserving mutual reductions between communicating tick automata and counter automata. As a consequence of the structure of our reductions, we show that channels and counters are mutually expressible, and similarly for emptiness tests and zero tests. This allows us to also obtain complexity results for communicating tick automata: We show that reachability in a system of communicating tick automata over a weakly-connected topology has the same complexity as reachability in Petri nets.[4]

*Related work.* Apart from [18], communication in a dense-time scenario has also been studied in [14,7,5]. In particular, [14] proposes timed message sequence charts as the semantics of communicating timed automata, and studies the scenario matching problem where timing constraints can be specified on local processes, later extended to also include send/receive pairs [7]. Communicating event-clock automata, a strict subclass of timed automata, are studied in [5] where, instead of considering the decidability frontier w.r.t. the communication topology, it is shown, among other results, that reachability is decidable for arbitrary topologies over existentially-bounded channels. A crucial difference w.r.t. our work is that we do not put any restriction on the channels, and we consider full timed automata. In a distributed setting, the model of global time we have chosen is not the only possible. In particular, [6] studies decidability of networks of (non-communicating) timed asynchronous automata in an alternative setting where each automaton has a local drift w.r.t. global time. In the discrete-time setting, we mention the work [17], which generalizes communicating tick automata to a loosely synchronous setting, where local times, though different, can differ at most by a given bound. While [17] shows decidability for a restricted two-processes topology, we characterize decidability for arbitrary topologies.

*Outline.* In Sec. 2 we introduce general notation; in particular, we define communicating timed processes, which allow us to uniformly model communication in both the discrete and dense time. In Sec. 3 we study the decidability and complexity for communicating tick automata (discrete time), while in Sec. 4 we deal with communicating timed automata (dense time). Finally, Sec. 5 ends the paper with future work. Full proofs are given in the appendix.

[4] The latter problem is known to be EXPSPACE-hard [20], and finding an upper bound is a long-standing open problem.

## 2 Communicating Timed Processes

A *labeled transition system* (LTS for short) is a tuple $\mathcal{A} = \langle S, S_I, S_F, A, \rightarrow \rangle$ where $S$ is a set of *states* with *initial states* $S_I \subseteq S$ and *final states* $S_F \subseteq S$, $A$ is a set of *actions*, and $\rightarrow \subseteq S \times A \times S$ is a *labeled transition relation*. For simplicity, we write $s \xrightarrow{a} s'$ in place of $(s, a, s') \in \rightarrow$. A *path* in $\mathcal{A}$ is an alternating sequence $\pi = s_0, a_1, s_1, \ldots, a_n, s_n$ of states $s_i \in S$ and actions $a_i \in A$ such that $s_{i-1} \xrightarrow{a_i} s_i$ for all $i \in \{1, \ldots, n\}$. We abuse notation and shortly denote $\pi$ by $s_0 \xrightarrow{a_1 \cdots a_n} s_n$. The word $a_1 \cdots a_n \in A^*$ is called the *trace* of $\pi$. A *run* is a path starting in an initial state ($s_0 \in S_I$) and ending in a final state ($s_n \in S_F$).

We consider systems that are composed of several processes interacting with each other in two ways. Firstly, they implicitly synchronize over the passing of time. Secondly, they explicitly communicate through the asynchronous exchange of messages. For the first point, we represent delays by actions in a given *delay domain* $\mathbb{D}$. Typically, the delay domain is a set of non-negative numbers when time is modeled quantitatively, or a finite set of abstract delays when time is modeled qualitatively. Formally, a *timed process* over $\mathbb{D}$ is a labeled transition system $\mathcal{A} = \langle S, S_I, S_F, A, \rightarrow \rangle$ such that $A \supseteq \mathbb{D}$. Actions in $A$ are either synchronous *delay actions* in $\mathbb{D}$, or *asynchronous actions* in $A \setminus \mathbb{D}$.

For the second point, we introduce fifo channels between processes. Formally, a communication *topology* is a triple $\mathcal{T} = \langle P, C, E \rangle$ where $\langle P, C \rangle$ is a directed graph comprising a finite set $P$ of *processes* and a set of communication *channels* $C \subseteq P \times P$, and, additionally, $E \subseteq C$ contains those channels that can be tested for emptiness. Thus, a channel $c \in C$ is a pair $(p, q)$, with the intended meaning that process $p$ can send messages to process $q$. For a process $p$, let $C[p] = \{q \mid (p, q) \in C\}$ be its set of outgoing channels, and let $C^{-1}[p] = \{q \mid (q, p) \in C\}$ be its set of incoming channels. Processes may send messages to outgoing channels, receive messages from incoming channels, as well as test emptiness of incoming channels (for testable channels). Formally, given a finite set $M$ of messages, the set of possible *communication actions* for process $p$ is $A^p_{\text{com}} = \{c!m \mid c \in C[p], m \in M\} \cup \{c?m \mid c \in C^{-1}[p], m \in M\} \cup \{c{==}\varepsilon \mid c \in E \cap C^{-1}[p]\}$. The set of all communication actions is $A_{\text{com}} = \bigcup_{p \in P} A^p_{\text{com}}$. While send actions ($c!m$) and receive actions ($c?m$) are customary, we introduce the extra test action ($c{==}\varepsilon$) to model the *urgent semantics* of [18] (cf. Appendix A.1).

**Definition 1.** *A* system of communicating timed processes *is a tuple* $\mathcal{S} = \langle \mathcal{T}, M, \mathbb{D}, (\mathcal{A}^p)_{p \in P} \rangle$ *where* $\mathcal{T} = \langle P, C, E \rangle$ *is a topology,* $M$ *is a finite set of* messages*,* $\mathbb{D}$ *is a delay domain, and, for each* $p \in P$*,* $\mathcal{A}^p = \langle S^p, S^p_I, S^p_F, A^p, \rightarrow^p \rangle$ *is a timed process over* $\mathbb{D}$ *such that* $A^p \cap A_{\text{com}} = A^p_{\text{com}}$*. Actions not in* $(\mathbb{D} \cup A_{\text{com}})$ *are called* internal actions.

States $s^p \in S^p$ are called *local states* of $p$, while a *global state* is a tuple of local states in $\prod_{p \in P} S^p$. We give the semantics of a system of communicating timed processes in terms of a global labeled transition system. The contents of each channel is represented as a finite word over the alphabet $M$. Processes move asynchronously, except for delay actions that occur simultaneously. Formally, the *semantics of a system of communicating timed processes* $\mathcal{S} = \langle \mathcal{T}, M, \mathbb{D}, (\mathcal{A}^p)_{p \in P} \rangle$ is the labeled transition system $[\![\mathcal{S}]\!] = \langle S, S_I, S_F, A, \rightarrow \rangle$ where $S = (\prod_{p \in P} S^p) \times (M^*)^C$, $S_I = (\prod_{p \in P} S_I^p) \times \{\lambda c \,.\, \varepsilon\}$, $S_F = (\prod_{p \in P} S_F^p) \times \{\lambda c \,.\, \varepsilon\}$, $A = \bigcup_{p \in P} A^p$, and there is a transition $(\boldsymbol{s}_1, w_1) \xrightarrow{a} (\boldsymbol{s}_2, w_2)$ under the following restrictions:

- if $a \in \mathbb{D}$, then $s_1^p \xrightarrow{a} s_2^p$ for all $p \in P$,
- if $a \notin \mathbb{D}$, then $s_1^p \xrightarrow{a} s_2^p$ for some $p \in P$, and $s_1^q = s_2^q$ for all $q \in P \setminus \{p\}$
  - if $a = c!m$, then $w_2(c) = w_1(c) \cdot m$ and $w_2(d) = w_1(d)$ for all $d \in C \setminus \{c\}$,
  - if $a = c?m$, then $m \cdot w_2(c) = w_1(c)$ and $w_2(d) = w_1(d)$ for all $d \in C \setminus \{c\}$,
  - if $a = (c \mathbin{==} \varepsilon)$, then $w_1(c) = \varepsilon$ and $w_1 = w_2$, and
  - if $a \notin A_{\text{com}}$, then $w_1 = w_2$.

To prevent confusion, states of $[\![\mathcal{S}]\!]$ will be called *configurations* in the remainder of the paper. Given a path $\pi$ in $[\![\mathcal{S}]\!]$, its *projection* to process $p$ is the path $\pi|_p$ in $\mathcal{A}^p$ obtained by projecting each transition of $\pi$ to process $p$ in the natural way.

The *reachability problem* asks, given a system of communicating timed processes $\mathcal{S}$, whether there exists a run in its semantics $[\![\mathcal{S}]\!]$. Note that we require all channels to be empty at the end of a run, which simplifies our constructions later by guaranteeing that every sent message is eventually received. (This is w.l.o.g. since reachability and control-state reachability are easily inter-reducible.) Two systems of communicating timed processes $\mathcal{S}$ and $\mathcal{S}'$ are said to be *equivalent* if $[\![\mathcal{S}]\!]$ has a run if and only if $[\![\mathcal{S}']\!]$ has a run.

**Definition 2.** *A* system of communicating tick automata *is a system of communicating timed processes* $\mathcal{S} = \langle \mathcal{T}, M, \mathbb{D}, (\mathcal{A}^p)_{p \in P} \rangle$ *such that* $\mathbb{D} = \{\tau\}$ *and each* $\mathcal{A}^p$ *is a* tick automaton*, i.e., a timed process over* $\mathbb{D}$ *with finitely many states and actions.*

Thus, tick automata communicate with actions in $A_{\text{com}}$ and, additionally, synchronize over the tick action $\tau$. This global synchronization makes communicating tick automata more expressive than CFSMs, in the sense that ticks can forbid re-orderings of communication actions that are legitimate without ticks (see Appendix A.2). Notice that there is only one tick symbol in $\mathbb{D}$: With two different ticks, reachability is already undecidable for the one channel topology $p \rightarrow q$ without emptiness test (see Appendix A.3).

## 3 Decidability of communicating tick automata

In this section, we study decidability and complexity of communicating tick automata. Our main technical tool consists of mutual reductions to/from counter automata, showing that, in the presence of tick actions, 1) each channel is equivalent to a counter, and 2) each emptiness test on a channel is equivalent to a zero test on the corresponding counter. This allows us to derive a complete characterization of decidable topologies, and also complexity results. We begin by defining communicating counter automata.

*Communicating counter automata.* A *counter automaton* is a classical Minsky machine $\mathcal{C} = \langle L, L_I, L_F, A, X, \Delta \rangle$ with finitely many locations $L$, initial locations $L_I \subseteq L$, final locations $L_F \subseteq L$, alphabet of actions $A$, finitely many counters in $X$, and transition rules $\Delta \subseteq L \times A \times L$. Operations on a counter $x \in X$ are $x$`++` (increment), $x$`--` (decrement) and $x$`==0` (zero test). Let $\mathtt{Op}(X)$ be the set of operations over counters in $X$. We require that $A \supseteq \mathtt{Op}(X)$. As usual, the semantics is given as a labelled transition system $[\![\mathcal{C}]\!] = \langle S, S_I, S_F, A, \rightarrow \rangle$ where $S = L \times \mathbb{N}^X$, $S_I = L_I \times \{\lambda x.0\}$, $S_F = L_F \times \{\lambda x.0\}$, and the transition relation $\rightarrow$ is defined as usual. Notice that acceptance is with zero counters.

A *system of communicating counter automata* is a system of communicating timed processes $\mathcal{S} = \langle \mathcal{T}, M, \mathbb{D}, ([\![\mathcal{C}^p]\!])_{p \in P} \rangle$ such that $\mathbb{D} = \emptyset$ and each $\mathcal{C}^p$ is a counter automaton. By Definition 1, this entails that each counter automaton performs communicating actions in $A^p_{com}$. Moreover, since the delay domain is empty, they can only interact through the asynchronous exchange of messages.

*From tick automata to counter automata.* Let $\mathcal{S}$ be a system of communicating tick automata over an arbitrary (i.e., possibly cyclic) weakly-connected[5] topology. We build an equivalent system of communicating counter automata $\mathcal{S}'$ over the same topology. Processes in $\mathcal{S}'$ are completely asynchronous, i.e., with empty delay domain.

Intuitively, we implement synchronization on the delay action $\tau$ in $\mathcal{S}$ by communication in $\mathcal{S}'$. We introduce a new type of message, also called $\tau$, which is sent in broadcast by all processes in $\mathcal{S}'$ each time there is a synchronizing tick action in $\mathcal{S}$. Since communication is by its nature asynchronous, we allow the sender and the receiver to be momentarily desynchronized during the computation. However, we restrict the desynchronization to be asymmetric: The receiver is allowed to be "ahead" of the sender (w.r.t. number of ticks performed), but never the other way around. This ensures causality between transmissions and receptions, by forbidding that a message is received before it is sent.

To keep track of the exact amount of desynchronization between sender and receiver (as a difference in number of ticks), we introduce counters in $\mathcal{S}'$: We endow each process $p$ with a non-negative counter $\mathtt{x}^p_c$ for each channel $c \in C^{-1}[p]$ from which $p$ is allowed to receive. The value of counter $\mathtt{x}^p_c$ measures the difference in number of ticks $\tau$ between $p$ and the corresponding sender along $c$. Whenever a process $p$ performs a synchronizing tick action $\tau$ in $\mathcal{S}$, in $\mathcal{S}'$ it sends a message $\tau$ in broadcast onto all outgoing channels; at the same time, all its counters $\mathtt{x}^p_c$ are incremented, recording that $p$, as a receiver process, is one more step ahead of its corresponding senders. When one such $\tau$-message is received by a process $q$ in $\mathcal{S}'$ along channel $c$, the corresponding counter $\mathtt{x}^q_c$ is decremented; similarly, this records that the sender process along $c$ is getting one step closer to the receiver process $q$. The topology needs to be weakly-connected for the correct propagation of $\tau$'s.

[5] A topology $\mathcal{T}$ is *weakly-connected* if, for every two processes, there is an undirected path between them.

While proper ordering of receptions and transmissions is ensured by non-negativeness of counters, testing emptiness of the channel is more difficult: In fact, a receiver, which in general is ahead of the sender, might see the channel as empty at one point (thus the test is positive), but then the sender might later (i.e., after performing some tick) send some message, and the earlier test should actually have failed (false positive). We avoid this difficulty by enforcing that the receiver $q$ is synchronized with the corresponding sender along channel $c$ on emptiness tests, by adding to the test action $c==\varepsilon$ by $q$ a zero test $\mathtt{x}_c^q\mathtt{==0}$.

Formally, let $\mathcal{S} = \langle \mathcal{T}, M, \mathbb{D}, (\mathcal{A}^p)_{p\in P}\rangle$ with $\mathbb{D} = \{\tau\}$ be a system of communicating tick automata over topology $\mathcal{T} = \langle P, C, E\rangle$, where, for each $p \in P$, $\mathcal{A}^p = \langle L^p, L^p_I, L^p_F, A^p, \to^p\rangle$ is a tick automaton, i.e., $\tau \in A^p$. We define the system of communicating counter automata $\mathcal{S}' = \langle \mathcal{T}, M', \mathbb{D}', ([\![\mathcal{C}^p]\!])_{p\in P}\rangle$, *over the same topology* $\mathcal{T}$ *as* $\mathcal{S}$, s.t. $M' = M \cup \{\tau\}$, $\mathbb{D}' = \emptyset$, and, for every process $p \in P$, we have a counter automaton $\mathcal{C}^p$, which is defined as follows: $\mathcal{C}^p = \langle L^p, L^p_I, L^p_F, B^p, \mathtt{X}^p, \Delta^p\rangle$, i.e., control locations $L^p$ in $\mathcal{C}^p$ are the same as locations in the corresponding tick automaton $\mathcal{A}^p$ (and also initial/final locations), and counters are those in $\mathtt{X}^p = \{\mathtt{x}^p_c \mid c \in C^{-1}[p]\}$. For simplifying the definition of transitions, we allow sequences of actions instead of just one action—these can be clearly implemented by introducing more intermediate states. Thus, transitions in $\mathcal{C}^p$ are defined as follows:

- Let $\ell \xrightarrow{\tau} \ell'$ be a transition in $\mathcal{A}^p$, and assume that outgoing channels of $p$ are those in $C[p] = \{c_0, \dots, c_h\}$, and that counters in $\mathtt{X}^p$ are those in $\{x_0, \dots, x_k\}$. Then, $\ell \xrightarrow{c_0!\tau;\dots;c_h!\tau;x_0\mathtt{++};\dots;x_k\mathtt{++}} \ell'$ is a transition in $\mathcal{C}^p$.
- For every $\ell \in L^p$ and input channel $c \in C^{-1}[p]$, there is a transition $\ell \xrightarrow{c?\tau;\mathtt{x}^p_c\mathtt{--}} \ell$ in $\mathcal{C}^p$.
- If $\ell \xrightarrow{c==\varepsilon} \ell'$ is a transition in $\mathcal{A}^p$, then $\ell \xrightarrow{\mathtt{x}^p_c\mathtt{==0};c==\varepsilon} \ell'$ is a transition in $\mathcal{C}^p$.
- Every other transition $\ell \xrightarrow{a} \ell'$ in $\mathcal{A}^p$ is also a transition in $\mathcal{C}^p$.

The action alphabet of $\mathcal{C}^p$ is thus $B^p = (A^p \setminus \{\tau\}) \cup \{c?\tau \mid c \in C^{-1}[p]\} \cup \{c!\tau \mid c \in C[p]\}$; in particular, $\tau$ is no longer an action, but a message that can be sent and received. We show that $\mathcal{S}$ and $\mathcal{S}'$ are equivalent, obtaining the following result.

**Proposition 1.** *Let* $\mathcal{T}$ *be a weakly-connected topology with* $\alpha$ *channels, of which* $\beta$ *can be tested for emptiness. For every system of communicating tick automata* $\mathcal{S}$ *with topology* $\mathcal{T}$*, we can produce, in linear time, an equivalent system of communicating counter automata* $\mathcal{S}'$ *with the same topology* $\mathcal{T}$*, containing* $\alpha$ *counters, of which* $\beta$ *can be tested for zero.*

While the proposition above holds for arbitrary weakly-connected topologies, it yields counter automata *with channels*, which are undecidable in general. To avoid undecidability due to communication, we need to forbid cycles (either directed or undirected) in the topology. It has been shown that, on polytrees[6], runs of communicating processes (even infinite-state) can be rescheduled as to satisfy the so-called *eagerness* requirement, where each transmission is *immediately* followed by the matching reception [16]. Their argument holds also in the presence of emptiness tests, since an eager run cannot disable $c == \varepsilon$ transitions (eager runs can only make the channels empty more often). Thus, by restricting to eager runs, communication behaves just as a rendezvous synchronization, and we obtain a global counter automaton by taking the product of all component counter automata.

[6] A *polytree* is a weakly-connected graph with neither directed, nor undirected cycles.

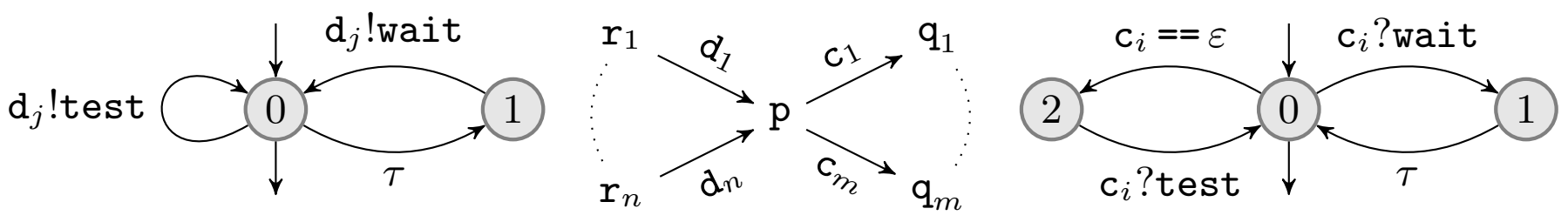

Fig. 1: Simulation of a counter automaton by a system of communicating tick automata: Tick automata for $\mathtt{r}_j$ (left) and $\mathtt{q}_i$ (right), Topology (middle).

**Theorem 1.** *For every polytree topology $\mathcal{T}$ with $\alpha$ channels, of which $\beta$ can be tested for emptiness, the reachability problem for systems of communicating tick automata with topology $\mathcal{T}$ is reducible, in linear time, to the reachability problem for products of (non-communicating) counter automata, with overall $\alpha$ counters, of which $\beta$ can be tested for zero.*

*From counter automata to tick automata.* We reduce the reachability problem for (non-communicating) counter automata to the reachability problem for systems of communicating tick automata with star topology. Formally, a topology $\mathcal{T} = \langle P, C, E \rangle$ is called a *star topology* if there exist two disjoint subsets $Q, R$ of $P$ and a process $p$ in $P \setminus (Q \cup R)$ such that $P = \{p\} \cup Q \cup R$ and $C = (R \times \{p\}) \cup (\{p\} \times Q)$. The idea is to simulate each counter with a separate channel, thus the number of counters fixes the number of channels in $\mathcal{T}$. However, our reduction is uniform in the sense that it works independently of the exact arrangement of channels in $\mathcal{T}$, which we take *not* to be under our control. W.l.o.g., we consider counter automata where all actions are counter operations (i.e., $\Delta \subseteq L \times \mathtt{Op}(X) \times L$).

For the remainder of this section, we consider an arbitrary star topology $\mathcal{T} = \langle P, C, E \rangle$ with set of processes $P = \{\mathtt{p}, \mathtt{q}_1, \ldots, \mathtt{q}_m, \mathtt{r}_1, \ldots, \mathtt{r}_n\}$, where $m, n \in \mathbb{N}$, and set of channels $C = \{\mathtt{p}\} \times \{\mathtt{q}_1, \ldots, \mathtt{q}_m\} \cup \{\mathtt{r}_1, \ldots, \mathtt{r}_n\} \times \{\mathtt{p}\}$ and $E = C$. This topology is depicted in Figure 1 (middle). Note that we allow the limit cases $m = 0$ and $n = 0$. To simplify the presentation, we introduce shorter notations for the channels of this topology: we define $\mathtt{c}_i = (\mathtt{p}, \mathtt{q}_i)$ and $\mathtt{d}_j = (\mathtt{r}_j, \mathtt{p})$ for every $i \in \{1, \ldots, m\}$ and $j \in \{1, \ldots, n\}$.

Let $\mathcal{C} = \langle L, L_I, L_F, X \cup Y, \Delta \rangle$ be a counter automaton with $m + n$ counters, namely $X = \{\mathtt{x}_1, \ldots, \mathtt{x}_m\}$ and $Y = \{\mathtt{y}_1, \ldots, \mathtt{y}_n\}$. The counters are split into $X$ and $Y$ to reflect the star topology $\mathcal{T}$, which is a priori given. We build, from $\mathcal{C}$, an equivalent system of communicating tick automata $\mathcal{S}$ with topology $\mathcal{T}$. Basically, the process $\mathtt{p}$ simulates the control-flow graph of the counter automaton, and the counters $\mathtt{x}_i$ and $\mathtt{y}_j$ are simulated by the channels $\mathtt{c}_i$ and $\mathtt{d}_j$, respectively. In order to define $\mathcal{S}$, we need to provide its message alphabet and its tick automata, one for each process $p$ in $P$. The message alphabet is $M = \{\mathtt{wait}, \mathtt{test}\}$. Actions performed by processes in $P$ are either communication actions or the delay action $\tau$. Processes $\mathtt{r}_j$'s are assigned the tick automaton of Figure 1 (left), and processes $\mathtt{q}_i$'s are assigned the tick automaton of Figure 1 (right). Intuitively, communications on $\mathtt{wait}$ messages are loosely synchronized using the $\tau$ actions in $\mathtt{q}_i$ and $\mathtt{r}_j$, so that $\mathtt{p}$ can control the rate of their reception and transmission.

We now present the tick automaton $\mathcal{A}^{\mathtt{p}}$. As mentioned above, the control-flow graph of $\mathcal{C}$ is preserved by $\mathcal{A}^{\mathtt{p}}$, so we only need to translate counter operations of $\mathcal{C}$ by communication actions and $\tau$ actions. Each counter operation of $\mathcal{C}$ is simulated by a finite sequence of actions in $\Sigma^{\mathtt{p}}$. To simplify the presentation, we directly label transitions of $\mathcal{A}^{\mathtt{p}}$ by words in $(\Sigma^{\mathtt{p}})^*$. The encoding of counter operations is given by the mapping $\eta$ from $\mathtt{Op}(X \cup Y)$ to $(\Sigma^{\mathtt{p}})^*$ defined as follows:

$$\begin{array}{rlrl}
\eta(\mathtt{x}_i\mathtt{++}) = & \mathtt{c}_i!\mathtt{wait} & \eta(\mathtt{x}_i\mathtt{--}) = & (\mathtt{c}_h!\mathtt{wait})_{1\le h\le m, h\ne i}\cdot\tau\cdot(\mathtt{d}_k?\mathtt{wait})_{1\le k\le n} \\
\eta(\mathtt{y}_j\mathtt{--}) = & \mathtt{d}_j?\mathtt{wait} & \eta(\mathtt{y}_j\mathtt{++}) = & (\mathtt{c}_h!\mathtt{wait})_{1\le h\le m}\cdot\tau\cdot(\mathtt{d}_k?\mathtt{wait})_{1\le k\le n, k\ne j} \\
\eta(\mathtt{x}_i\mathtt{==0}) = & \mathtt{c}_i!\mathtt{test} & \eta(\mathtt{y}_j\mathtt{==0}) = & (\mathtt{d}_j == \varepsilon)\cdot(\mathtt{d}_j?\mathtt{test})
\end{array}$$

where $i \in \{1, \ldots, m\}$ and $j \in \{1, \ldots, n\}$. We obtain $\mathcal{A}^{\mathtt{p}}$ from $\mathcal{C}$ by replacing each counter operation by its encoding. Observe that these replacements require the addition of a set $S^{\mathtt{p}}_{\diamond}$ of fresh intermediate states to implement sequences of actions. Formally, $\mathcal{A}^{\mathtt{p}}$ is the tick automaton $\mathcal{A}^{\mathtt{p}} = \langle L \cup S^{\mathtt{p}}_{\diamond}, L_I, L_F, \Sigma^{\mathtt{p}}, \{\ell \xrightarrow{\eta(\mathtt{op})} \ell' \mid (\ell, \mathtt{op}, \ell') \in \Delta\}\rangle$. This completes the definition of the system of communicating tick automata $\mathcal{S} = \langle \mathcal{T}, M, \{\tau\}, (\mathcal{A}^p)_{p\in P}\rangle$.

A formal proof that $[\![\mathcal{C}]\!]$ has a run if and only if $[\![\mathcal{S}]\!]$ has a run is provided in Appendix C.3. Here, we only explain the main ideas behind this simulation of $\mathcal{C}$ by $\mathcal{S}$. The number of $\mathtt{wait}$ messages in channels $\mathtt{c}_i$ and $\mathtt{d}_j$ encodes the value of counters $\mathtt{x}_i$ and $\mathtt{y}_j$, respectively. So, incrementing $\mathtt{x}_i$ amounts to sending $\mathtt{wait}$ in $\mathtt{c}_i$, and decrementing $\mathtt{y}_j$ amounts to receiving $\mathtt{wait}$ from $\mathtt{d}_j$. Both actions can be performed by $\mathtt{p}$. Decrementing $\mathtt{x}_i$ is more involved, since $\mathtt{p}$ cannot receive from the channel $\mathtt{c}_i$. Instead, $\mathtt{p}$ performs a $\tau$ action in order to force a $\tau$ action in $\mathtt{q}_i$, hence, a receive of $\mathtt{wait}$ by $\mathtt{q}_i$. But all other processes also perform the $\tau$ action, so $\mathtt{p}$ compensates (see the definition of $\eta(\mathtt{x}_i\mathtt{--})$) in order to preserve the number of $\mathtt{wait}$ messages in the other channels. The simulation of $\mathtt{y}_j\mathtt{++}$ by $\eta(\mathtt{y}_j\mathtt{++})$ is similar. Let us now look at zero test operations. When $\mathtt{p}$ simulates $\mathtt{x}_i\mathtt{==0}$, it simply sends $\mathtt{test}$ in the channel $\mathtt{c}_i$. This message is eventually received by $\mathtt{q}_i$ since all channels must be empty at the end of the simulation. The construction guarantees that the first receive action of $\mathtt{q}_i$ after the send action $\mathtt{c}_i!\mathtt{test}$ of $\mathtt{p}$ is the matching receive $\mathtt{c}_i?\mathtt{test}$. This means, in particular, that the channel is empty when $\mathtt{p}$ sends $\mathtt{test}$ in $\mathtt{c}_i$. The same device is used to simulate a zero test of $\mathtt{y}_j$, except that the roles of $\mathtt{p}$ and its peer (here, $\mathtt{r}_j$) are reversed. Clearly, channels that need to be tested for emptiness are those encoding counters that are tested for zero. We obtain the following theorem.

**Theorem 2.** *Let $\mathcal{T}$ be an a priori given star topology with $\alpha$ channels, of which $\beta$ can be tested for emptiness. The reachability problem for (non-communicating)* *counter automata with $\alpha$ counters, of which $\beta$ can be tested for zero, is reducible, in linear time, to the reachability problem for systems of communicating tick automata with topology $\mathcal{T}$.*

*Decidability and complexity results for communicating tick automata.* Thanks to the mutual reductions to/from counter automata developed previously, we may now completely characterize which topologies (not necessarily weakly-connected) have a decidable reachability problem, depending on exactly which channels can be tested for emptiness. Intuitively, decidability still holds even in the presence of multiple emptiness tests, provided that each test appear in a different weakly-connected component.

**Theorem 3 (Decidability).** *Given a topology $\mathcal{T}$, the reachability problem for systems of communicating tick automata with topology $\mathcal{T}$ is decidable if and only if $\mathcal{T}$ is a polyforest*[7] *containing at most one testable channel in each weakly-connected component.*

*Proof.* For one direction, assume that the reachability problem for systems of communicating tick automata with topology $\mathcal{T}$ is decidable. The topology $\mathcal{T}$ is necessarily a polyforest, since the reachability problem is undecidable for non-polyforest topologies even without ticks [21,19]. Suppose that $\mathcal{T}$ contains a weakly-connected component with (at least) two channels that can be tested for emptiness. By an immediate extension of Theorem 2 to account for the undirected path between these two channels, we can reduce the reachability problem for two-counter automata to the reachability problem for systems of communicating tick automata with topology $\mathcal{T}$. Since the former is undecidable, each weakly-connected component in $\mathcal{T}$ contains at most one testable channel.

For the other direction, assume that $\mathcal{T}$ is a polyforest with at most one testable channel in each weakly-connected component, and let $\mathcal{S}$ be a system of communicating tick automata with topology $\mathcal{T}$. Thus, $\mathcal{S}$ can be decomposed into a disjoint union of independent systems $\mathcal{S}_0, \mathcal{S}_1, \ldots, \mathcal{S}_n$, where each $\mathcal{S}_k$ has an undirected tree topology containing exactly one testable channel. But we need to ensure that the $\mathcal{S}_k$'s perform the same number of ticks. By (the construction leading to) Theorem 1, each $\mathcal{S}_k$ can be transformed into an equivalent counter automaton $\mathcal{C}_k$ (by taking the product over all processes in $\mathcal{S}_k$), where exactly one counter, let us call it $\mathtt{x}_k$, can be tested for zero. We may suppose, w.l.o.g., that the counters of $\mathcal{C}_0, \ldots, \mathcal{C}_n$ are disjoint. Moreover, $\mathcal{C}_k$ can maintain, in an extra counter $\mathtt{y}_k$, the number of ticks performed by $\mathcal{S}_k$. We compose the counter machines $\mathcal{C}_0, \ldots, \mathcal{C}_n$ sequentially, and check, at the end, that $\mathtt{y}_0 = \cdots = \mathtt{y}_n$. Since all counters must be zero in final configurations, this check can be performed by adding, on the final state, a loop decrementing all the $\mathtt{y}_k$'s simultaneously. The construction guarantees that the resulting global counter machine $\mathcal{C}$ is equivalent to $\mathcal{S}$. However, $\mathcal{C}$ contains zero tests on many counters: $\mathtt{x}_0, \ldots, \mathtt{x}_n$. Fortunately, these counters are used one after the other, and they are zero at the beginning and at the end. So we may re-use $\mathtt{x}_0$ in place of $\mathtt{x}_1, \ldots, \mathtt{x}_n$. We only need to check that $\mathtt{x}_0$ is zero when switching from $\mathcal{C}_k$ to $\mathcal{C}_{k+1}$. Thus, we have reduced the reachability problem for systems of communicating tick automata with topology $\mathcal{T}$ to the reachability problem for counter automata with zero tests on only one counter. As the latter is decidable [22,10], the former is decidable, too.

[7] A topology $\mathcal{T}$ is a *polyforest* if it is a directed acyclic graph with no undirected cycle.

When no test is allowed, we obtain a simple characterization of the complexity for polyforest topologies. A topology $\mathcal{T} = \langle P, C, E\rangle$ is *test-free* if $E = \emptyset$.

**Corollary 1 (Complexity).** *The reachability problem for systems of communicating tick automata with test-free polyforest topologies has the same complexity as the reachability problem for counter automata without zero tests (equivalently, Petri nets).*

*Remark 1.* Even though global synchronization makes communicating tick automata more expressive than CFSMs, our characterization shows that they are decidable for exactly the same topologies (polyforest). However, while reachability for CFSMs is PSPACE-complete, systems of communicating tick automata are equivalent to Petri nets, for which reachability is EXPSPACE-hard [20] (the upper bound being a long-standing open problem).

## 4 Decidability of communicating timed automata

In this section, we consider communicating timed automata, which are communicating timed processes synchronizing over the dense delay domain $\mathbb{D} = \mathbb{R}_{\geq 0}$. We extend the decidability results for tick automata of Section 3 to the case of timed automata. To this end, we present mutual, topology-preserving reductions between communicating tick automata and communicating timed automata. We first introduce the latter model.

*Communicating timed automata.* A *timed automaton* $\mathcal{B} = \langle L, L_I, L_F, X, \Sigma, \Delta\rangle$ is defined by a finite set of locations $L$ with initial locations $L_I \subseteq L$ and final locations $L_F \subseteq L$, a finite set of *clocks* $X$, a finite alphabet $\Sigma$ and a finite set $\Delta$ of transitions rules $(\ell, \sigma, g, R, \ell')$ where $\ell, \ell' \in L$, $\sigma \in \Sigma$, the guard $g$ is a conjunction of constraints $x \# c$ for $x \in X$, $\# \in \{<, \leq, =, \geq, >\}$ and $c \in \mathbb{N}$, and $R \subseteq X$ is a set of clocks to reset.

The semantics of $\mathcal{B}$ is given by the timed process $[\![\mathcal{B}]\!] = \langle S, S_I, S_F, A, \to\rangle$, where $S = L \times \mathbb{R}^X_{\geq 0}$, $S_I = L_I \times \{\lambda x.\, 0\}$, $S_F = L_F \times \{\lambda x.\, 0\}$, $A = \Sigma \cup \mathbb{R}_{\geq 0}$ is the set of actions, and there is a transition $(\ell, v) \xrightarrow{d} (\ell, v')$ if $d \in \mathbb{R}_{\geq 0}$ and $v'(x) = v(x) + d$ for every clock $x$, and $(\ell, v) \xrightarrow{\sigma} (\ell', v')$ if there exists a rule $(\ell, \sigma, g, R, \ell') \in \Delta$ such that $g$ is satisfied by $v$ (defined in the natural way) and $v'(x) = 0$ when $x \in R$, $v'(x) = v(x)$ otherwise. We decorate a path $(\ell_0, u_0) \xrightarrow{a_0, t_0} (\ell_1, u_1) \xrightarrow{a_1, t_1} \cdots (a_n, u_n)$ in $[\![\mathcal{B}]\!]$ with additional *timestamps* $t_i = \sum\{a_j \mid j = 0, \ldots, i-1 \text{ and } a_j \in \mathbb{R}_{\geq 0}\}$. Note that we require cloks to be zero on accepting runs, which simplifies a construction later.[8] W.l.o.g. we do not consider location invariants in timed automata as they can be encoded in the guards; reachability is preserved since acceptance with zero cloks forbids the elapse of time upon entering the last location of an accepting run.

[8] It can be implemented by duplicating final locations, and by resetting all clocks when entering the new final locations.

A *system of communicating timed automata* is a system of communicating timed processes $\mathcal{S} = \langle \mathcal{T}, M, \mathbb{R}_{\geq 0}, (\llbracket \mathcal{B}^p \rrbracket)_{p \in P} \rangle$ where each $\mathcal{B}^p$ is a timed automaton. Note that each timed automaton has access only to its local clocks. By Definition 1, each timed automaton performs communicating actions in $A^p_{com}$ and synchronizes with all the other processes over delay actions in $\mathbb{R}_{\geq 0}$.

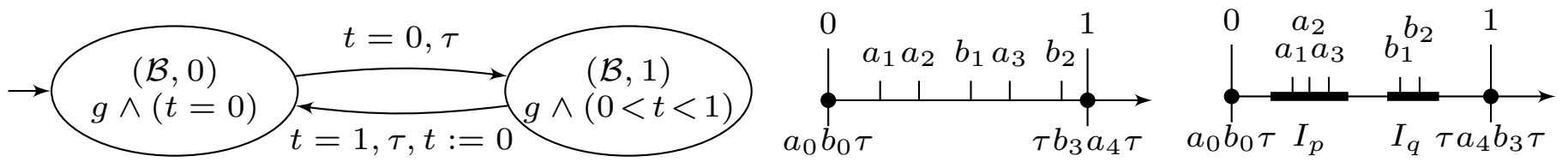

Fig. 2: From timed to tick automata: instrumentation of a timed automaton $\mathcal{B}$ with $\tau$-transitions (left), addition of $\tau$'s along a run (middle) and rescheduling of a run (right).

*From timed automata to tick automata.* On test-free acyclic topologies, we show a topology-preserving reduction from communicating timed to communicating tick automata. We insist on a reduction that only manipulates processes locally, thus preserving the topology. The absence of emptiness tests on the channels enables such a modular construction.

Naïvely, one would just apply the classical region construction to each process [8]. However, while this preserves local reachability properties, it does not respect the global synchronization between different processes. While quantitative synchronization cannot be obtained by locally removing dense time, a qualitative synchronization suffices in our setting. We require that all processes are either at the same integer date $k \in \mathbb{N}$, or in the same open interval $(k, k+1)$. This suffices because, at integer dates (in fact, at any time-point), any interleaving is allowed, and, in intervals $(k, k+1)$, we can reschedule all processes s.t., for every channel $c = (p, q)$, all actions of $p$ occur before all actions of $q$ (cf. the Rescheduling Lemma below). The latter property ensures the causality between transmissions and receptions.

Qualitative synchronization is achieved by forcing each automaton $\mathcal{B}^p$ to perform a synchronizing tick action $\tau$ at each date $k$ and at each interval $(k, k+1)$. See Figure 2 on the left, where $\mathcal{B}^p$ is split into two copies $(\mathcal{B}^p, 0)$ and $(\mathcal{B}^p, 1)$: Actions occurring on integer dates $k$ are performed in $(\mathcal{B}^p, 0)$, and those in $(k, k+1)$ happen in $(\mathcal{B}^p, 1)$. This is ensured by adding a new clock $\mathtt{t}$ and $\tau$-transitions that switch from one mode to the other. Formally, the *$\tau$-instrumentation* of $\mathcal{B} = \langle L, L_I, L_F, X, \Sigma, \Delta \rangle$ is the timed automaton $\mathsf{Instr}(\mathcal{B}, \tau) = \langle L \times \{0,1\}, L_I \times \{1\}, F \times \{0,1\}, X \cup \{\mathtt{t}\}, \Sigma \cup \{\tau\}, \Delta' \rangle$, where $\mathtt{t} \notin X$ and $\Delta'$ is defined by: $(\ell, 0) \xrightarrow{a, (g \wedge \mathtt{t}=0), R} (\ell', 0)$ and $(\ell, 1) \xrightarrow{a, (g \wedge 0 < \mathtt{t} < 1), R} (\ell', 1)$ for all rules $\ell \xrightarrow{a, g, R} \ell'$ in $\Delta$, and $(\ell, 0) \xrightarrow{\tau, \mathtt{t}=0, \emptyset} (\ell, 1)$ and $(\ell, 1) \xrightarrow{\tau, \mathtt{t}=1, \{\mathtt{t}\}} (\ell, 0)$ for all locations $\ell \in L$.

Finally, we obtain an equivalent system of tick automata by applying the exponential region construction to each instrumented process.

**Theorem 4.** *Let $\mathcal{T}$ be a test-free acyclic topology. For every system of communicating timed automata $\mathcal{S} = \langle \mathcal{T}, M, \mathbb{R}_{\geq 0}, (\llbracket \mathcal{B}^p \rrbracket)_{p \in P} \rangle$ with topology $\mathcal{T}$, we can produce, in exponential time, an equivalent system of communicating tick automata $\mathcal{S}' = \langle \mathcal{T}, M, \{\tau\}, (\mathcal{A}^p)_{p \in P} \rangle$ over the same topology $\mathcal{T}$, where the tick automaton $\mathcal{A}^p$ is obtained by applying the region graph construction to $\mathsf{Instr}(\mathcal{B}^p, \tau)$.*

One direction of the equivalence between $\mathcal{S}$ and $\mathcal{S}'$ is immediate, since every run in $\mathcal{S}$ induces a run in $\mathcal{S}'$ by just inserting $\tau$ actions in the right position. For the other direction, let $\rho'$ be a run of $\mathcal{S}'$, and we show how to build a corresponding run $\rho$ of $\mathcal{S}$. We have to schedule all the actions in $\rho'$ on timestamps that are consistent with the guards in $\mathcal{S}$ and that preserve dependencies between transmissions and receptions of messages. Consider a channel $c = (p, q)$ without emptiness test. If $p$ and $q$ are untimed processes, it is always possible to first schedule transmissions of $p$, and then receptions of $q$. The Rescheduling Lemma below ensures the same for timed processes. This is depicted in Figure 2 in the middle (before rescheduling) and on the right (after rescheduling) where the $a$'s are emissions of $p$ and the $b$'s are receptions of $q$.

**Lemma 1 (Rescheduling Lemma)** *Let $\mathcal{B}$ be a timed automaton, and $I \subseteq (0, 1)$ an open interval. Then, every run of $\mathcal{B}$ $(\ell_0, v_0) \xrightarrow{a_0, t_0} \cdots (\ell_n, v_n)$ can be rescheduled such that integral timestamps $t_i \in \mathbb{N}$ are kept the same, and non-integral timestamps $t_i \in (k, k+1)$ belong to $k + I$.*

Intuitively, the lemma above allows us to restrict non-integer timestamps in $(k, k+1)$ to occur in a predefined sub-interval $I+k$. Let us first see how this helps in constructing $\rho'$. To each process $p$, we associate an open interval $I_p \subseteq (0, 1)$, such that, for every channel $(p, q)$, $I_p$ and $I_q$ are disjoint, and $I_p$ comes before $I_q$. This is always possible on acyclic topologies. Then, all actions of process $p$ in $(k, k+1)$ are rescheduled to occur in $k + I_p$ (according to the Recheduling Lemma), which ensures causality between transmissions and receptions. Finally, the $\tau$ actions added by instrumentation tell, for each action performed by process $p$ in $\rho'$, whether it should be scheduled at an integer date $k$, or in $k + I_p$.

*Remark 2.* We show in Appendix D.2 that our reduction is incorrect in the presence of emptiness tests. We also show that there are essential difficulties in rescheduling senders and receivers in fixed intervals, as emptiness tests introduce a sort of circular dependency and seem to require unboundedly many intervals.

We now comment about the correctness of the Rescheduling Lemma (proved in Appendix D.1). Resets and guards in a timed automaton allow to enforce minimal and/or maximal delays between timestamps on a path. Since clocks are compared to integers only, it suffices to just distinguish between integral and non-integral dates. While for closed guards like $x \leq 1$ a non-integral time-point $t \in (0, 1)$ would suffice to represent all non-integral dates, to accommodate open guards like $x < 1$ we need a dense interval $I \subseteq (0, 1)$. The following characterization of decidable test-free topologies follows from Theorems 3 and 4.

**Theorem 5 (Decidability).** *Given a test-free topology $\mathcal{T}$, the reachability problem for systems of communicating timed automata with topology $\mathcal{T}$ is decidable if and only if $\mathcal{T}$ is a polyforest.*

*Remark 3.* While the reachability problem is known to be decidable for a system of two communicating timed automata with only one channel and emptiness test [18], that proof does not preserve the topology and it looks hardly adaptable to arbitrary polyforest topologies.

*From tick automata to timed automata.* Given a system of communicating tick automata $\mathcal{S}$, we produce an equivalent system of communicating timed automata $\mathcal{S}'$, over the same topology. The synchronization on $\tau$'s is easily simulated using clocks in $\mathcal{S}'$ by ensuring that all the processes elapse 1 time unit exactly when they (synchronously) perform a $\tau$ in $\mathcal{S}$. Thus, every run in $\mathcal{S}$ has a corresponding run in $\mathcal{S}'$. For the converse to hold, we have to make sure that for every run of $\mathcal{S}'$, all the processes perform the same number of $\tau$'s on the corresponding run of $\mathcal{S}$. This ensured since we require clocks to be zero at the end of accepting runs, thus preventing time to elapse on final locations.

The simple topology $p \to q \to r$ is known to be undecidable when both channels can be tested for emptiness [18]. Thanks to Theorem 3, we obtain generalized undecidability for every weakly-connected topology containing at least two testable channels.

**Theorem 6 (Undecidability).** *Given a weakly-connected topology $\mathcal{T}$ with two testable channels, the reachability problem for systems of communicating timed automata with topology $\mathcal{T}$ is undecidable.*

## 5 Conclusions and future work

We have studied the decidability and complexity of communicating timed processes. In discrete time, we give a complete characterization of decidable topologies with emptiness tests, as well as a tight connection with Petri nets in the test-free case. In dense time, we prove decidability for polyforest test-free topologies, and we generalize the undecidability results of [18] to arbitrary weakly-connected topologies containing two testable channels. We leave open whether one can obtain, in the presence of emptiness tests, the same characterization as in discrete time. We conjecture that this is possible, although the techniques used here do not seem to easily extend to deal with emptiness tests. Finally, as another direction for future work one can study richer models where processes are allowed to send timestamps or clocks along channels, in the spirit of [1].

*Acknowledgements.* We thank Jérôme Leroux, Anca Muscholl, and Igor Walukiewicz for helpful discussions.

## References

1. P. A. Abdulla, M. F. Atig, and J. Cederberg. Timed lossy channel systems. In *FSTTCS*, LIPIcs, 2012. To appear.
2. P. A. Abdulla, M. F. Atig, and J. Stenman. Dense-timed pushdown automata. In *LICS*, pages 35–44, 2012.
3. P. A. Abdulla and B. Jonsson. Verifying programs with unreliable channels. *Information and Computation*, 127(2):91–101, 1996.
4. P. A. Abdulla and A. Nylén. Timed Petri Nets and BQOs. In *ICATPN*, volume 2075 of *LNCS*, pages 53–70, 2001.
5. S. Akshay, B. Bollig, and P. Gastin. Automata and logics for timed message sequence charts. In *FSTTCS*, volume 4855 of *LNCS*, pages 290–302, 2007.
6. S. Akshay, B. Bollig, P. Gastin, M. Mukund, and K. N. Kumar. Distributed timed automata with independently evolving clocks. In *CONCUR*, volume 5201 of *LNCS*, pages 82–97, 2008.
7. S. Akshay, P. Gastin, M. Mukund, and K. N. Kumar. Model checking time-constrained scenario-based specifications. In *FSTTCS*, volume 8 of *LIPIcs*, pages 204–215, 2010.
8. R. Alur and D. Dill. A theory of timed automata. *TCS*, 126(2):183–235, 1994.
9. B. Bérard, F. Cassez, S. Haddad, D. Lime, and O. H. Roux. Comparison of Different Semantics for Time Petri Nets. In *ATVA*, volume 3707 of *LNCS*, pages 293–307. 2005.
10. R. Bonnet. The reachability problem for vector addition system with one zero-test. In *MFCS*, volume 6907 of *LNCS*, pages 145–157, 2011.
11. D. Brand and P. Zafiropulo. On communicating finite-state machines. *J. ACM*, 30(2):323–342, 1983.
12. G. Cécé and A. Finkel. Verification of programs with half-duplex communication. *Information and Computation*, 202(2):166–190, 2005.
13. P. Chambart and Ph. Schnoebelen. Mixing lossy and perfect fifo channels. In *CONCUR*, volume 5201 of *LNCS*, pages 340–355, 2008.
14. P. Chandrasekaran and M. Mukund. Matching scenarios with timing constraints. In *FORMATS*, volume 4202 of *LNCS*, pages 98–112, 2006.
15. H. Gruber, M. Holzer, A. Kiehn, and B. König. On timed automata with discrete time - structural and language theoretical characterization. In *DLT*, volume 3572 of *LNCS*, pages 272–283, 2005.
16. A. Heußner, J. Leroux, A. Muscholl, and G. Sutre. Reachability analysis of communicating pushdown systems. *Logical Methods in Comp. Sci.*, 8(3:23):1–20, 2012.
17. O. H. Ibarra, Z. Dang, and P. S. Pietro. Verification in loosely synchronous queue-connected discrete timed automata. *Theor. Comput. Sci.*, 290(3):1713–1735, 2003.
18. P. Krcál and W. Yi. Communicating timed automata: The more synchronous, the more difficult to verify. In *CAV*, volume 4144 of *LNCS*, pages 249–262, 2006.
19. S. La Torre, P. Madhusudan, and G. Parlato. Context-bounded analysis of concurrent queue systems. In *TACAS*, volume 4963 of *LNCS*, pages 299–314, 2008.
20. R. J. Lipton. *The Reachability Problem Requires Exponential Space.* Department of Computer Science, Yale University, 1976.
21. J. K. Pachl. Reachability problems for communicating finite state machines. Research Report CS-82-12, University of Waterloo, May 1982.
22. K. Reinhardt. Reachability in petri nets with inhibitor arcs. *ENTCS*, 223:239–264, 2008.

# A On Communicating Timed Processes

## A.1 Modeling urgency with emptiness test

We show how the urgent semantics of [18] can be modelled with a test for empty channel. In the *urgent semantics* for receive actions of [18], if a message can be received by a process, then internal actions are disabled (while other communication and delay actions are still enabled). In our model, instead of defining a separate urgent semantics, we introduce the extra test action $c\,\texttt{==}\,\varepsilon$, which allows us to discuss more precisely where in the topology is the urgent semantics (i.e., test action) used. Below, we show how to implement the urgent semantics of [18] with the test action.

We need to ensure that internal actions of control states where also a receive action $c?m$ is available can be executed only if $m$ cannot be received from $c$. In turn, this can only happen iff either $c$ is empty, or it is not empty and the message in front of the channel is $m' \neq m$. Let $M(\ell) = \{m \mid \ell \xrightarrow{c?m} \ell'\}$ be the set of messages that can be read from a given control location $\ell$. For the second condition, we modify the automaton with a standard construction to store into its finite control the first message $m'$ that can be received (if any), and check that $m \notin M(\ell)$ before the internal action can be executed. For the first condition, in the case no message $m'$ is in the local buffer, the internal action is preceded by a test action $c\,\texttt{==}\,\varepsilon$ (by introducing an intermediate state).

## A.2 On the power of ticks

Consider the topology with two processes $q$ and $r$ and a channel from $q$ to $r$ (that cannot be tested for emptiness). Formally, this topology is the triple $\mathcal{U} = \langle \{q, r\}, \{(q, r)\}, \emptyset \rangle$. It is known that every CFSM with topology $\mathcal{U}$ is existentially 1-bounded, i.e., each run can be re-ordered into a run where the channel always contains at most one message [21,16]. However, this property doesn't hold for systems of communicating tick automata.

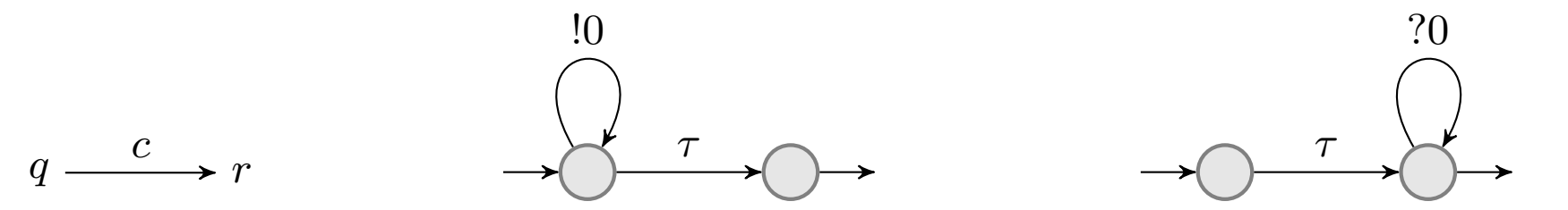

(a) Topology

(b) Tick automaton for process $q$

(c) Tick automaton for process $r$

Fig. 3: A simple system of communicating tick automata that is *not* existentially-bounded.

Consider the example depicted in Figure 3. Because of the global synchronization enforced by the tick action $\tau$, the first reception necessarily occurs after the last transmission. Hence, this example is not existentially-bounded: for every bound $B \in \mathbb{N}$, there exists a run with no $B$-bounded re-ordering. This shows that systems of communicating tick automata are more expressive than CFSM. Alternatively, from a language viewpoint, the trace language of this example is $\{(!0)^n\tau(?0)^n \mid n \in \mathbb{N}\}$. However, no CFSM (with topology $\mathcal{U}$) has the same trace language (where $\tau$ would be an internal action).

### A.3 Undecidability of multi-tick automata

One could consider a more expressive model where communicating tick automata can synchronize over a finite set of distinct tick actions $\{\tau_1, \tau_2, \ldots, \tau_k\}$, instead of just one tick $\tau$. However, in the simplest non-trivial topology $\mathcal{T}' = \langle\{q, r\}, \{(q, r)\}, \emptyset\rangle$ (no emptiness tests) with two processes $q, r$ and a channel from $q$ to $r$ (as in Figure 4a), reachability becomes undecidable already with $k = 2$ tick actions. In fact, a perfect channel automaton $\mathcal{S} = \langle\langle\{p\}, \{(p, p)\}, \emptyset\rangle, M, \emptyset, \{\mathcal{A}^p\}\rangle$ (for which reachability is undecidable [11]) can be simulated by topology $\mathcal{T}'$ above. Without loss of generality, assume $M = \{0, 1\}$. $\mathcal{S}$ can be simulated by two communicating finite-state automata (i.e., CFSMs) $\mathcal{S}' = \langle\mathcal{T}', M, \mathbb{D}, \{\mathcal{A}^q, \mathcal{A}^r\}\rangle$ over topology $\mathcal{T}' = \langle\{q, r\}, \{(q, r)\}, \emptyset\rangle$ as above, and where $\mathbb{D} = \{\tau_0, \tau_1\}$, $\mathcal{A}^r$ is shown in Figure 4b, and $\mathcal{A}^q$ is defined as follows. Let $c$ be the channel $(q, r)$. The send actions $!m$ of $p$ are seamlessly performed by $q$ as $c!m$. Since $q$ (unlike $p$) cannot directly read from the channel (only $r$ can), for simulating a receive action $?m$ of $p$, $m \in \{0, 1\}$, $q$ performs the corresponding tick action $\tau_m$ in order to force process $r$ to read the correct message $m$ on its behalf.

**Theorem 7.** *Let $\mathcal{T}$ be a topology with at least one channel. Then, the reachability problem for communicating multi-tick automata with at least two distinct tick actions and with topology $\mathcal{T}$ is undecidable.*

## B Proofs of Section 3

### B.1 From tick automata to counter automata

For simplifying the presentation of the proof, we allow broadcast transmission of $\tau$-messages via actions of the form $C[p]!\tau$ and global increment actions $\mathtt{X}^p$`++` on the set of counters $\mathtt{X}^p$. Thus, the first case in the definition of transitions in $\mathcal{C}^p$ is as follows:

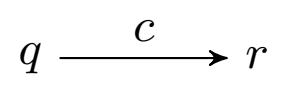

(a) Topology

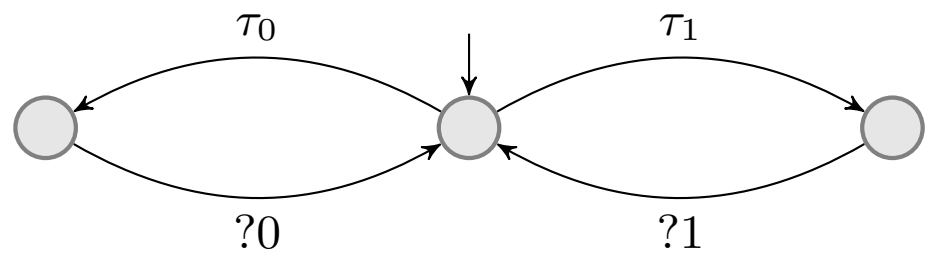

(b) Multi-tick automaton for process $r$

Fig. 4: Simulation of a perfect channel automaton by a 2 tick automaton.

– If $\ell \xrightarrow{\tau} \ell'$ is a transition in $\mathcal{A}^p$, then $\ell \xrightarrow{C[p]!\tau;\mathtt{x}^p\mathtt{++}} \ell'$ is a transition in $\mathcal{C}^p$.

**Proposition 1.** *Let $\mathcal{T}$ be a weakly-connected topology with $\alpha$ channels, of which $\beta$ can be tested for emptiness. For every system of communicating tick automata $\mathcal{S}$ with topology $\mathcal{T}$, we can produce, in linear time, an equivalent system of communicating counter automata $\mathcal{S}'$ with the same topology $\mathcal{T}$, containing $\alpha$ counters, of which $\beta$ can be tested for zero.*

*Proof.* Given $\mathcal{S} = \langle \mathcal{T}, M, \{\tau\}, (\mathcal{A}^p)_{p\in P}\rangle$, let $\mathcal{S}' = \langle \mathcal{T}, M \cup \{\tau\}, \emptyset, ([\![\mathcal{C}^p]\!])_{p\in P}\rangle$ be as defined in Section 3. We show that a run in $\mathcal{S}$ induces a run in $\mathcal{S}'$, and vice versa.

For the first direction, assume there exists a run $\pi$ in $\mathcal{S}$. We obtain a run $\pi'$ in $\mathcal{S}'$ by a simple manipulation of $\pi$. First, all transitions in $\pi$ different from $\tau$ and $c \mathtt{==} \varepsilon$ can be taken as they are in $\pi'$. Second, if there is a $\tau$ transition in $\pi$, then it is replaced in $\mathcal{S}'$ by any interleaving of transitions in $\{\ell \xrightarrow{C[p]!\tau;\mathtt{x}^p\mathtt{++}} \ell' \mid p \in P\}$; after this sequence of transitions, control locations in $\pi$ and $\pi'$ match again for each process. Moreover, the matching receive transitions $\ell'' \xrightarrow{c?\tau;\mathtt{x}_c^p\mathtt{--}} \ell''$ are introduced later as soon as they can be fired (control locations do not change), so that $\tau$ messages do not get stuck in the channels preventing other non-$\tau$ messages from being received. Notice that these transitions are enabled since each counter $\mathtt{x}_c^p$ is incremented when the $\tau$ is performed (by process $p$), and decremented when a message $\tau$ sent at the same time from some other process $q$ along $c$ is subsequently received by $p$. Thus, the counter $\mathtt{x}_c^p$ is always $> 0$ when a $\tau$ is received, and the transitions above are always enabled. Moreover, when a channel $c = (p, q)$ is empty, the sender $p$ and the receiver $q$ have performed the same number of $\tau$'s (since the $\tau$'s are sent in contiguous blocks), and $\mathtt{x}_c^q$ is zero. Finally, a transition $\ell \xrightarrow{c\mathtt{==}\varepsilon} \ell'$ in $\mathcal{S}$ is translated to a transition $\ell \xrightarrow{\mathtt{x}_c^p\mathtt{==0};c\mathtt{==}\varepsilon} \ell'$ in $\mathcal{S}'$, which can be fired since, by the observation above, $\mathtt{x}_c^p$ is zero when $c$ is empty.

For the other direction, let $\pi_0$ be a run in $\mathcal{S}'$. We reorder transitions in $\pi_0$ in order to obtain another run $\pi_1$ in $\mathcal{S}'$ in which processes are synchronized on $\tau$'s. Then, $\pi_1$ is directly mapped to a run $\pi_2$ in $\mathcal{S}$ by replacing transitions in $\mathcal{S}'$ with the matching transitions in $\mathcal{S}$.

*From $\pi_0$ to $\pi_1$.* We now explain how to translate from $\pi_0$ to $\pi_1$. Since $\mathcal{S}'$ is a completely asynchronous system, we can view $\pi_0$ as a sequence of transitions $\pi_0 = t_0, t_1, \ldots, t_n$, where each transition $t_i$ is fired by some process $p_i$. Assume that such a transition has the form $t_i = (\ell_i, v_i) \xrightarrow{b_i} (\ell_i', v_i')$, where $\ell_i, \ell_i'$ are locations of $p_i$, $v_i, v_i'$ are valuations for $p_i$'s counters, and $b_i$ is an action in $B^{p_i}$. Moreover, for each process $p$, let $\pi_0|_p$ be the projection of $\pi_0$ containing only transitions belonging to process $p_i = p$. The idea is to decorate transitions $t_i$ in $\pi_0$ with an integral *timestamp* $k_i(p) \geq 0$ counting how many $\tau$'s have been sent so far by process $p$ (on any fixed channel). Formally, $k_i(p)$ is the number of transitions $t_j$ in $\pi_0|_p$ s.t. $i < j$ (i.e., excluding $t_i$ itself) and $b_j = C[p]!\tau$.

Let $\pi'_0 = (t_0, k_0, w_0), (t_1, k_1, w_1), \ldots, (t_n, k_n, w_n)$ be the decorated path, where, additionally, channel valuations $w_i$'s are added recording the global contents of the channels before transition $t_i$ is fired. Finally, let $\#\tau_i(c)$ be the number of messages $\tau$ in $w_i(c)$. A few observations are in order:

- At the beginning, $k_0(p) = 0$ for every process $p$.
- For each $p$, the sequence $k_0(p), k_1(p), \ldots, k_n(p)$ is non-decreasing.
- For each channel $c = (p, q)$, the receiver process $q$ has received, at step $i$, $k_i(q) - v_i(\mathtt{x}_c^q) \geq 0$ $\tau$-messages.[9] Consequently, there are $\#\tau_i(c) = k_i(p) - (k_i(q) - v_i(\mathtt{x}_c^q)) \geq 0$ $\tau$'s left on channel $c$. When $c$ is empty,
$$k_i(p) = k_i(q) - v_i(\mathtt{x}_c^q) \tag{1}$$
- At the end, $k_n(p) = k_n(q)$ for every processes $p$ and $q$ (since channels are empty and counters are zero).

However, while timestamps are locally non-decreasing, they are not necessarily globally non-decreasing. Having globally non-decreasing timestamps is necessary to show that the processes can be correctly synchronized on $\tau$'s. We produce another run $\pi_1$ starting from $\pi'_0$, where timestamps are not only locally non-decreasing, but also globally non-decreasing. To do so, we show that transitions in $\pi'_0$ can be swapped when the timestamp decreases (necessarily along different processes). Formally, we swap adjacent transitions

$$(t_i, k_i, w_i), (t_{i+1}, k_{i+1}, w_{i+1}) \quad \text{whenever} \quad k_i(p_i) > k_{i+1}(p_{i+1}) \tag{†}$$

In general, we say that a pair of transitions $(t_i, t_j)$ with $i < j$ is *offending* iff $k_i(p_i) > k_j(p_j)$; we aim at a new run $\pi_1$ with no offending transitions. Notice that in a path with no offending transitions, once a process broadcasts a $\tau$ (by simulating a tick action), then it is blocked until all other processes have done the same.

The difficulty in swapping offending transitions is that, in general, transitions might have dependencies between each other, and dependent transitions cannot be swapped. We analyse the dependencies that can theoretically arise, and we argue that offending transitions cannot be dependent, and thus they are swappable. There are three kinds of dependencies for a pair of transitions $(t_i, t_j)$, $i < j$:

1. *Locality:* $t_i$ and $t_j$ belong to the same process $p_i = p_j$.
2. *Send/Receive:* $t_i$ is a send on a channel $c$ and $t_j$ is the matching receive.
3. *Test/Send:* $t_i$ is an emptiness test $b_i = c \mathtt{==} \varepsilon$ on $c$, and $t_j$ is the first send on $c$ since $t_i$. Formally, $b_j = c!m$, and for every $i < k < j$ and $m' \in M$, $b_k \neq c!m'$ (thus, $w_i(c) = w_{i+1}(c) = \cdots = w_j(c) = \varepsilon$).

We argue that offending transitions cannot be dependent, therefore we can swap all transitions as in (†) above. Clearly, when no more transitions can be swapped, we have globally non-decreasing timestamps, and the swapping process terminates since the total number of offending pairs decreases at each step. Thus, let $(t_i, k_i, w_i), (t_{i+1}, k_{i+1}, w_{i+1})$ be two adjacent offending transitions, i.e., $k_i(p_i) > k_{i+1}(p_{i+1})$. We show that they are not dependent for each one of the cases above:

1. *Locality:* Clearly, $t_i, t_{i+1}$ belong to different processes ($p_i \neq p_{i+1}$) since timestamps are locally non-decreasing. Thus, there is no locality dependency.
2. *Send/Receive:* Since the transitions are offending, $k_i(p_i) > k_{i+1}(p_{i+1})$, thus $p_i$ has sent more $\tau$'s than $p_{i+1}$ has done. By how counters are updated (being always non-negative), $p_{i+1}$ cannot receive more $\tau$'s from $p_i$ than it has sent himself. Therefore, $p_i$ has sent more $\tau$'s than $p_{i+1}$ has *received*, thus there are $\tau$'s still in the channel. Formally, $\#\tau_{i+1}(c) > 0$ by Equation 1 (since, by local non-decreasingness, $k_i(p_i) \leq k_{i+1}(p_i)$). Therefore, the message sent from $p_i$ is not in front of the channel and cannot be received by $p_{i+1}$, and there is no Send/Receive dependency.
3. *Test/Send:* Since $t_i$ is an emptiness test $b_i = c \texttt{==} \varepsilon$ on $c$, $w_i(c) = w_{i+1}(c) = \varepsilon$. By construction, process $p_i$ has previously checked that counter $\mathtt{x}_c^{p_i}$ is zero. Since the counter can only be modified by $p_i$, $v_i(\mathtt{x}_c^{p_i}) = 0$, and, since the counter does not change in the next step, also $v_{i+1}(\mathtt{x}_c^{p_i}) = 0$. Therefore, by Equation 1, $k_{i+1}(p_{i+1}) = k_{i+1}(p_i)$. Since $k_i(p_i) = k_{i+1}(p_i)$ (no new $\tau$'s have been performed by $p_i$), we get $k_i(p_i) = k_{i+1}(p_{i+1})$, which is a contradiction since transitions are offending. Thus, there is no Test/Send dependency.

[9] In fact, counter $\mathtt{x}_c^q$ is incremented $k_i(q)$ times, and decremented each time a $\tau$-message is received.

*From $\pi_1$ to $\pi_2$.* We have thus built a non-offending sequence of $\mathcal{S}'$-transitions $\pi_1 = (t'_0, k'_0, w'_0), (t'_1, k'_1, w'_1), \ldots, (t'_n, k'_n, w'_n)$. The former can be transformed into a sequence of $\mathcal{S}$-transitions $\pi_2 = (e_0, m_0), (e_1, m_1), \ldots, (e_n, m_n)$ (decorated with channel contents $m_i$) by discarding the timestamp annotations $k'_i$, by removing $\tau$'s from channels (i.e., $m_i(c)$ equals $w'_i(c)$ without $\tau$'s), and by inverting transition-wise the construction, as follows: For every transition $t'_i = (\ell_i, v_i) \xrightarrow{b_i} (\ell'_i, v'_i)$ in $\mathcal{S}'$, we define the transition $e_i = \ell_i \xrightarrow{a_i} \ell'_i$ in $\mathcal{S}$ by case analysis (we set $a_i$ equal to the special symbol $\epsilon$ when the transition is to be removed):

- Transmitting a $\tau$-message becomes a tick action $\tau$: If $b_i = C[p]!\tau$, then $a_i = \tau$.
- Receiving of $\tau$'s disappears: If $b_i = c?\tau$, then $a_i = \epsilon$.
- Counter operations disappear: If $b_i \in \mathtt{Op}'(\mathtt{X}^p)$ then, $a_i = \epsilon$.
- Every other action stays unchanged, i.e., $a_i = b_i$ for every other $b_i$'s.
  In particular, for tests of channel emptiness, if $b_i = c \texttt{==} \varepsilon$, then $a_i = c \texttt{==} \varepsilon$. Since $w'_i(c) = \varepsilon$, then $m_i(c) = \varepsilon$ and $e_i$ can be fired.

Let $k$ be the total number of processes $p$'s. Since $\pi_1$ was non-offending, tick actions $\tau$ in $\pi_2$ occur in blocks of length exactly $k$, one for each $p$. Therefore, the sequence of transitions $\pi_2$ can be interpreted in a path of $\mathcal{S}$ where the processes synchronize on $\tau$'s.

### B.2 Complexity

**Corollary 1 (Complexity).** *The reachability problem for systems of communicating tick automata with test-free polyforest topologies has the same complexity* *as the reachability problem for counter automata without zero tests (equivalently, Petri nets).*

*Proof.* The lower bound follows immediately from Theorem 2. For the upper bound, we use the same construction as in the proof of Theorem 3. However, each component $\mathcal{C}_i$ in that construction was derived as a product of counter automata (cf. Theorem 1), which would introduce an exponential blow-up in the number of locations. We avoid the blowup by a standard construction replacing each location in each process in $\mathcal{C}_i$ by a (1-bounded) counter, and adding a finite control to simulate the transitions of $\mathcal{C}_i$.

### C.3 From counter automata to tick automata

We formally prove, in this appendix subsection, the simulation of counter automata by systems of communicating tick automata with star topology. This simulation was presented, informally, in Section 3. We refer the reader to this section for the definition of the constructed system of communicating tick automata $\mathcal{S}$ with star topology $\mathcal{T}$.

Recall that the set $\boldsymbol{S}$ of global states of $\mathcal{S}$ is the cartesian product of its sets of local states, i.e., $\boldsymbol{S} = \prod_{p \in P} S^p$. To simplify notation, global states of $\mathcal{S}$ will also be denoted by triples $(t, \boldsymbol{u}, \boldsymbol{v})$ where $t \in S^{\mathtt{p}}$, $\boldsymbol{u} \in \prod_{i=1}^{m} S^{\mathtt{q}_i}$ and $\boldsymbol{v} \in \prod_{j=1}^{n} S^{\mathtt{r}_j}$. We write $\mathbf{0}$ for the vector $(0, \ldots, 0)$ and $\mathbf{1}$ for the vector $(1, \ldots, 1)$. For every valuation $v \in \mathbb{N}^{X \cup Y}$, we define the *encoding* $\eta(v) \in (M^*)^C$ of $v$ by $\eta(v)(\mathtt{c}_i) = \mathtt{wait}^{v(\mathtt{x}_i)}$ and $\eta(v)(\mathtt{d}_j) = \mathtt{wait}^{v(\mathtt{y}_j)}$. The following lemma shows that every transition in $[\![\mathcal{C}]\!]$ can be simulated by a path of $[\![\mathcal{S}]\!]$.

**Lemma 1.** *For every transition* $(\ell, v) \xrightarrow{\mathtt{op}} (\ell', v')$ *of* $[\![\mathcal{C}]\!]$*, there exists a path from* $((\ell, \mathbf{0}, \mathbf{0}), \eta(v))$ *to* $((\ell', \mathbf{0}, \mathbf{0}), \eta(v'))$ *in* $[\![\mathcal{S}]\!]$*.*

*Proof.* Consider a transition $(\ell, v) \xrightarrow{\mathtt{op}} (\ell', v')$ of $[\![\mathcal{C}]\!]$. To simplify notation, we define $w = \eta(v)$ and $w' = \eta(v')$. A couple of intermediate states in $S^{\mathtt{p}}_{\diamond}$ are sometimes needed to decompose paths. We will simply denote them by $\diamond_1$ and $\diamond_2$. We consider six cases, depending on the counter operation $\mathtt{op}$.

- $\mathtt{x}_i$++. It holds that $v'(\mathtt{x}_i) = v(\mathtt{x}_i) + 1$ and $v'(x) = v(x)$ for all $x \in X \cup Y$ with $x \neq \mathtt{x}_i$. Hence, $w'(\mathtt{c}_i) = w(\mathtt{c}_i) \cdot \mathtt{wait}$ and $w'(c) = w(c)$ for all $c \in C$ with $c \neq \mathtt{c}_i$. By construction, $[\![\mathcal{S}]\!]$ contains the following transition:
$$((\ell, \mathbf{0}, \mathbf{0}), w) \xrightarrow{\mathtt{c}_i!\mathtt{wait}} ((\ell', \mathbf{0}, \mathbf{0}), w')$$
- $\mathtt{y}_j$--. It holds that $v(\mathtt{y}_j) = v'(\mathtt{y}_j) + 1$ and $v'(x) = v(x)$ for all $x \in X \cup Y$ with $x \neq \mathtt{y}_j$. Hence, $w(\mathtt{d}_j) = \mathtt{wait} \cdot w'(\mathtt{d}_j)$ and $w'(c) = w(c)$ for all $c \in C$ with $c \neq \mathtt{d}_j$. By construction, $[\![\mathcal{S}]\!]$ contains the following transition:
$$((\ell, \mathbf{0}, \mathbf{0}), w) \xrightarrow{\mathtt{d}_j?\mathtt{wait}} ((\ell', \mathbf{0}, \mathbf{0}), w')$$

– $\mathtt{x}_i$--. It holds that $v(\mathtt{x}_i) = v'(\mathtt{x}_i) + 1$ and $v'(x) = v(x)$ for all $x \in X \cup Y$ with $x \neq \mathtt{x}_i$. Hence, $w(\mathtt{c}_i) = \mathtt{wait} \cdot w'(\mathtt{c}_i)$ and $w'(c) = w(c)$ for all $c \in C$ with $c \neq \mathtt{c}_i$. Furthermore, $w(c) \in \{\mathtt{wait}\}^*$ for all $c \in C$. By construction, $[\![\mathcal{S}]\!]$ contains the following path:

$$((\ell, \mathbf{0}, \mathbf{0}), w) \xrightarrow{(\mathtt{c}_h!\mathtt{wait})_{1 \leq h \leq m, h \neq i}} \cdot \xrightarrow{(\mathtt{c}_i?\mathtt{wait})_{1 \leq i \leq m}} ((\diamond_1, \mathbf{1}, \mathbf{0}), w') \xrightarrow{\tau}$$
$$((\diamond_2, \mathbf{0}, \mathbf{1}), w') \xrightarrow{(\mathtt{d}_j!\mathtt{wait})_{1 \leq j \leq n}} \cdot \xrightarrow{(\mathtt{d}_k?\mathtt{wait})_{1 \leq k \leq n}} ((\ell', \mathbf{0}, \mathbf{0}), w')$$

– $\mathtt{y}_j$++. It holds that $v'(\mathtt{y}_j) = v(\mathtt{y}_j) + 1$ and $v'(x) = v(x)$ for all $x \in X \cup Y$ with $x \neq \mathtt{y}_j$. Hence, $w'(\mathtt{d}_j) = w(\mathtt{d}_j) \cdot \mathtt{wait}$ and $w'(c) = w(c)$ for all $c \in C$ with $c \neq \mathtt{d}_j$. Furthermore, $w(c) \in \{\mathtt{wait}\}^*$ for all $c \in C$. By construction, $[\![\mathcal{S}]\!]$ contains the following path:

$$((\ell, \mathbf{0}, \mathbf{0}), w) \xrightarrow{(\mathtt{c}_h!\mathtt{wait})_{1 \leq h \leq m}} \cdot \xrightarrow{(\mathtt{c}_i?\mathtt{wait})_{1 \leq i \leq m}} ((\diamond_1, \mathbf{1}, \mathbf{0}), w) \xrightarrow{\tau}$$
$$((\diamond_2, \mathbf{0}, \mathbf{1}), w) \xrightarrow{(\mathtt{d}_k!\mathtt{wait})_{1 \leq k \leq n}} \cdot \xrightarrow{(\mathtt{d}_k?\mathtt{wait})_{1 \leq k \leq n, k \neq j}} ((\ell', \mathbf{0}, \mathbf{0}), w')$$

– $\mathtt{x}_i$==0. It holds that $v = v'$ and $v(\mathtt{x}_i) = v'(\mathtt{x}_i) = 0$. Hence, $w = w'$ and $w(\mathtt{c}_i) = w'(\mathtt{c}_i) = \varepsilon$. By construction, $[\![\mathcal{S}]\!]$ contains the following path:

$$((\ell, \mathbf{0}, \mathbf{0}), w) \xrightarrow{\mathtt{c}_i == \varepsilon} \cdot \xrightarrow{\mathtt{c}_i!\mathtt{test}} \cdot \xrightarrow{\mathtt{c}_i?\mathtt{test}} ((\ell', \mathbf{0}, \mathbf{0}), w')$$

– $\mathtt{y}_j$==0. It holds that $v = v'$ and $v(\mathtt{y}_j) = v'(\mathtt{y}_j) = 0$. Hence, $w = w'$ and $w(\mathtt{d}_j) = w'(\mathtt{d}_j) = \varepsilon$. By construction, $[\![\mathcal{S}]\!]$ contains the following path:

$$((\ell, \mathbf{0}, \mathbf{0}), w) \xrightarrow{\mathtt{d}_j == \varepsilon} \cdot \xrightarrow{\mathtt{d}_j!\mathtt{test}} \cdot \xrightarrow{\mathtt{d}_j?\mathtt{test}} ((\ell', \mathbf{0}, \mathbf{0}), w')$$

We get, in all cases, that there is a path from $((\ell, \mathbf{0}, \mathbf{0}), w)$ to $((\ell', \mathbf{0}, \mathbf{0}), w')$ in $[\![\mathcal{S}]\!]$.

For the reverse direction, we show that paths of $[\![\mathcal{S}]\!]$ encoding a single counter operation correspond to transitions of $[\![\mathcal{C}]\!]$. This correspondence is expressed as follows. For every $\boldsymbol{s} \in \boldsymbol{S}$ and $w \in (M^*)^C$, we define the *decoding* $\delta(\boldsymbol{s}, w) \in \mathbb{N}^{X \cup Y}$ of $(\boldsymbol{s}, w)$ by

$$\delta(\boldsymbol{s}, w)(\mathtt{x}_i) \;=\; |w(\mathtt{c}_i)|_{\mathtt{wait}} + (s^{\mathtt{q}_i} \bmod 2) \qquad \text{and} \qquad \delta(\boldsymbol{s}, w)(\mathtt{y}_j) \;=\; |w(\mathtt{d}_j)|_{\mathtt{wait}} + s^{\mathtt{r}_j}$$

where $|u|_{\mathtt{wait}}$ denotes the number of occurences of wait in a word $u \in M^*$. Since p is the process controlling the simulation of the counter machine, the decoding should remain constant along transitions that do not involve p. It is routinely checked that this property holds.

*Remark 4.* It holds that $\delta(\boldsymbol{s}, w) = \delta(\boldsymbol{s}', w')$ for every transition $(\boldsymbol{s}, w) \xrightarrow{a} (\boldsymbol{s}', w')$ of $[\![\mathcal{S}]\!]$ such that $a \notin \Sigma^{\mathtt{p}} \cup \{\tau\}$.

**Lemma 2.** *For every operation $op \in Op(X \cup Y)$ and for every path $\pi = (\boldsymbol{s}, w) \xrightarrow{*} (\boldsymbol{s}', w')$ in $[\![\mathcal{S}]\!]$, $(\ell, \delta(\boldsymbol{s}, w)) \xrightarrow{op} (\ell', \delta(\boldsymbol{s}', w'))$ is a transition of $[\![\mathcal{C}]\!]$ if*

*1. the projection of* $\pi$ *on* $\mathtt{p}$ *is the extended transition* $\ell \xrightarrow{\eta(\mathtt{op})} \ell'$*, and*
*2. there exists* $\boldsymbol{s}'' \in \boldsymbol{S}$ *such that* $(\boldsymbol{s}', w') \xrightarrow{*} (\boldsymbol{s}'', \lambda c\,.\,\varepsilon)$ *in* $[\![\mathcal{S}]\!]$.

*Proof.* Consider a counter operation $\mathtt{op} \in \mathtt{Op}(X \cup Y)$ and a path $\pi = (\boldsymbol{s}, w) \xrightarrow{*} (\boldsymbol{s}', w')$ in $[\![\mathcal{S}]\!]$. Assume that both conditions of the lemma are satisfied. To simplify notation, define $v = \delta(\boldsymbol{s}, w)$ and $v' = \delta(\boldsymbol{s}', w')$. Let us show that $(\ell, v) \xrightarrow{\mathtt{op}} (\ell', v')$ is a transition of $[\![\mathcal{C}]\!]$. By assumption, $\mathcal{A}^{\mathtt{p}}$ contains the extended transition $\ell \xrightarrow{\eta(\mathtt{op})} \ell'$. Since $\eta$ is injective, we get that $(\ell, \mathtt{op}, \ell') \in \Delta$. It remains to prove that $v$ and $v'$ conform to the semantics of counter automata. We consider six cases, depending on the counter operation $\mathtt{op}$.

- $\mathtt{x}_i$++. The path $\pi$ may be written as $\pi = \chi_1 \cdot (\boldsymbol{s}_1, w_1) \xrightarrow{\mathtt{c}_i!\mathtt{wait}} (\boldsymbol{s}_2, w_2) \cdot \chi_2$. Since $\chi_1$ and $\chi_2$ do not involve $\mathtt{p}$, it holds that $v = \delta(\boldsymbol{s}, w) = \delta(\boldsymbol{s}_1, w_1)$ and $\delta(\boldsymbol{s}_2, w_2) = \delta(\boldsymbol{s}', w') = v'$. Hence, $v'(\mathtt{x}_i) = v(\mathtt{x}_i) + 1$ and $v'(x) = v(x)$ for all $x \in X \cup Y$ with $x \neq \mathtt{x}_i$.
- $\mathtt{y}_j$--. The path $\pi$ may be written as $\pi = \chi_1 \cdot (\boldsymbol{s}_1, w_1) \xrightarrow{\mathtt{d}_j?\mathtt{wait}} (\boldsymbol{s}_2, w_2) \cdot \chi_2$. By proceeding as above, we get that $v(\mathtt{y}_j) = v'(\mathtt{y}_j) + 1$ and $v'(x) = v(x)$ for all $x \in X \cup Y$ with $x \neq \mathtt{y}_j$.
- $\mathtt{x}_i$--. The path $\pi$ may be written as $\pi = \chi_1 \cdot (\boldsymbol{s}_1, w_1) \xrightarrow{\tau} (\boldsymbol{s}_2, w_2) \cdot \chi_2$. Observe that $\delta(\boldsymbol{s}_2, w_2)(x) = \delta(\boldsymbol{s}_1, w_1)(x) - 1$ and $\delta(\boldsymbol{s}_2, w_2)(y) = \delta(\boldsymbol{s}_1, w_1)(y) + 1$, for all $x \in X$ and $y \in Y$. The projection of $\chi_1$ and $\chi_2$ on $\mathtt{p}$ have trace $(\mathtt{c}_h!\mathtt{wait})_{1 \leq h \leq m, h \neq i}$ and $(\mathtt{d}_k?\mathtt{wait})_{1 \leq k \leq n}$, respectively. We derive that
  - $v'(\mathtt{x}_i) = \delta(\boldsymbol{s}_2, w_2)(\mathtt{x}_i) = \delta(\boldsymbol{s}_1, w_1)(\mathtt{x}_i) - 1 = v(\mathtt{x}_i) - 1$.
  - for all $x \in X$ with $x \neq \mathtt{x}_i$, $v'(x) = \delta(\boldsymbol{s}_2, w_2)(x) = \delta(\boldsymbol{s}_1, w_1)(x) - 1 = v(x)$.
  - for all $y \in Y$, $v'(y) = \delta(\boldsymbol{s}_2, w_2)(y) - 1 = \delta(\boldsymbol{s}_1, w_1)(y) = v(y)$.

  Hence, $v(\mathtt{x}_i) = v'(\mathtt{x}_i) + 1$ and $v'(x) = v(x)$ for all $x \in X \cup Y$ with $x \neq \mathtt{x}_i$.
- $\mathtt{y}_j$++. The path $\pi$ may be written as $\pi = \chi_1 \cdot (\boldsymbol{s}_1, w_1) \xrightarrow{\tau} (\boldsymbol{s}_2, w_2) \cdot \chi_2$. Again, $\delta(\boldsymbol{s}_2, w_2)(x) = \delta(\boldsymbol{s}_1, w_1)(x) - 1$ and $\delta(\boldsymbol{s}_2, w_2)(y) = \delta(\boldsymbol{s}_1, w_1)(y) + 1$, for all $x \in X$ and $y \in Y$. The projection of $\chi_1$ and $\chi_2$ on $\mathtt{p}$ have trace $(\mathtt{c}_h!\mathtt{wait})_{1 \leq h \leq m}$ and $(\mathtt{d}_k?\mathtt{wait})_{1 \leq k \leq n, k \neq j}$, respectively. By proceeding as above, we get that $v'(\mathtt{y}_j) = v(\mathtt{y}_j) + 1$ and $v'(x) = v(x)$ for all $x \in X \cup Y$ with $x \neq \mathtt{y}_j$.
- $\mathtt{x}_i$==0. The path $\pi$ may be written as $\pi = \chi_1 \cdot (\boldsymbol{s}_1, w_1) \xrightarrow{\mathtt{c}_i!\mathtt{test}} (\boldsymbol{s}_2, w_2) \cdot \chi_2$. Note that $\delta(\boldsymbol{s}_1, w_1) = \delta(\boldsymbol{s}_2, w_2)$. Since $\chi_1$ and $\chi_2$ do not involve $\mathtt{p}$, we obtain that $v = \delta(\boldsymbol{s}_1, w_1) = \delta(\boldsymbol{s}_2, w_2) = v'$. Let us show that $v(\mathtt{x}_i) = 0$. By assumption, it is possible to reach, from $(\boldsymbol{s}', w')$, a configuration with all channels empty. Therefore, there exists a path $(\boldsymbol{s}_1, w_1) \xrightarrow{\mathtt{c}_i!\mathtt{test}} (\boldsymbol{s}_2, w_2) \cdot \xi \cdot (\boldsymbol{s}_3, w_3) \xrightarrow{\mathtt{c}_i?\mathtt{test}} (\boldsymbol{s}_4, w_4)$ in $[\![\mathcal{S}]\!]$ such that its last action, $\mathtt{c}_i?\mathtt{test}$, is the matching receive of its first action, $\mathtt{c}_i!\mathtt{test}$. This means that $w_1(\mathtt{c}_i)$ is precisely the sequence of messages received from $\mathtt{c}_i$ in $\xi$. Observe that the channel $\mathtt{c}_i$ remains non-empty in $\xi$. Therefore, $\xi$ does not contain the action $\mathtt{c}_i \mathtt{==} \varepsilon$. By construction, this entails that the projection of $\xi$ on $\mathtt{q}_i$ is empty. It follows that $s_1^{\mathtt{q}_i} = s_3^{\mathtt{q}_i} = 2$. Moreover, since $\mathtt{q}_i$ is the receiver of $\mathtt{c}_i$ and $w_1(\mathtt{c}_i)$ is entirely received in $\xi$, we derive that $w_1(\mathtt{c}_i) = \varepsilon$. Hence, $v(\mathtt{x}_i) = \delta(\boldsymbol{s}_1, w_1)(\mathtt{x}_i) = 0$.

– $\mathtt{y}_j\mathtt{==0}$. The path $\pi$ may be written as $\pi = \chi_1 \cdot (\boldsymbol{s}_1, w_1) \xrightarrow{\mathtt{d}_j\,\mathtt{==}\,\varepsilon} (\boldsymbol{s}_2, w_2) \cdot \chi_2 \cdot (\boldsymbol{s}_3, w_3) \xrightarrow{\mathtt{d}_j\mathtt{?test}} (\boldsymbol{s}_4, w_4) \cdot \chi_3$. Note that $\delta(\boldsymbol{s}_1, w_1) = \delta(\boldsymbol{s}_2, w_2)$ and $\delta(\boldsymbol{s}_3, w_3) = \delta(\boldsymbol{s}_4, w_4)$. Since $\chi_1, \chi_2$ and $\chi_3$ do not involve $\mathtt{p}$, we obtain that $v = \delta(\boldsymbol{s}_1, w_1) = \cdots = \delta(\boldsymbol{s}_4, w_4) = v'$. Let us show that $v(\mathtt{y}_j) = 0$. Obviously, it holds that $w_1(\mathtt{d}_j) = w_2(\mathtt{d}_j) = \varepsilon$. Moreover, since $\chi_2$ does not contain any reception from $\mathtt{d}_j$, the first message sent to $\mathtt{d}_j$ in $\chi_2$ is $\mathtt{test}$, which entails that $s_1^{\mathtt{r}_j} = s_2^{\mathtt{r}_j} = 0$. Hence, $v(\mathtt{y}_j) = \delta(\boldsymbol{s}_1, w_1)(\mathtt{y}_j) = 0$.

We obtain, in all cases, that $v$ and $v'$ conform to the semantics of the counter operation $\mathtt{op}$. Since $(\ell, \mathtt{op}, \ell') \in \Delta$, we conclude that $(\ell, v) \xrightarrow{\mathtt{op}} (\ell', v')$ is a transition of $[\![\mathcal{C}]\!]$.

**Proposition 2.** *There exists a run in $[\![\mathcal{C}]\!]$ if and only if there exists a run in $[\![\mathcal{S}]\!]$.*

*Proof.* Consider a run $\rho = (\ell, v) \xrightarrow{*} (\ell', v')$ in $[\![\mathcal{C}]\!]$. By applying Lemma 1 to each transition of $\rho$, we obtain a path from $((\ell, \mathbf{0}, \mathbf{0}), \eta(v))$ to $((\ell', \mathbf{0}, \mathbf{0}), \eta(v'))$ in $[\![\mathcal{S}]\!]$. This path is a run since $((\ell, \mathbf{0}, \mathbf{0}), \eta(v))$ and $((\ell', \mathbf{0}, \mathbf{0}), \eta(v'))$ are initial and final configurations of $[\![\mathcal{S}]\!]$, respectively.

To prove the converse, pick a run $\rho = (\boldsymbol{s}, w) \xrightarrow{*} (\boldsymbol{s}', w')$ in $[\![\mathcal{S}]\!]$. The projection $\rho|_{\mathtt{p}}$ of $\rho$ on $\mathtt{p}$ is a path in $\mathcal{A}^{\mathtt{p}}$ starting and ending in $L$. Hence, $\rho|_{\mathtt{p}}$ may be written as a concatenation $\ell_0 \xrightarrow{\eta(\mathtt{op}_1)} \ell_1 \cdots \ell_{k-1} \xrightarrow{\eta(\mathtt{op}_k)} \ell_k$ of extended transitions. It follows that $\rho$ is a concatenation $\rho = \pi_1 \cdots \pi_k$ of paths $\pi_i$ in $[\![\mathcal{S}]\!]$ such that $\pi_i|_{\mathtt{p}} = \ell_{i-1} \xrightarrow{\eta(\mathtt{op}_i)} \ell_i$ for all $i \in \{1, \ldots, k\}$. Since $\rho$ ends in a configuration with all channels empty, each path $\pi_i$ satisfies the two conditions of Lemma 2. We obtain, by applying Lemma 2 to each $\pi_i$, a path from $(\ell, \delta(\boldsymbol{s}, w))$ to $(\ell', \delta(\boldsymbol{s}', w'))$ in $[\![\mathcal{C}]\!]$. This path is a run since $(\ell, \delta(\boldsymbol{s}, w))$ and $(\ell', \delta(\boldsymbol{s}', w'))$ are initial and final configurations of $[\![\mathcal{C}]\!]$, respectively.

# D Appendix of Section 4

## D.1 Proof of the Rescheduling Lemma

We first restate the Rescheduling Lemma.

**Lemma 2** *Let $\mathcal{B}$ be a timed automaton, and $I \subseteq (0, 1)$ an open interval. Then, every run of $\mathcal{B}$ $(\ell_0, v_0) \xrightarrow{a_0, t_0} \cdots (\ell_n, v_n)$ can be rescheduled such that integral timestamps $t_i \in \mathbb{N}$ are kept the same, and non-integral timestamps $t_i \in (k, k+1)$ are rescheduled in $k + I$.*

Let us first introduce notations and preliminary definitions. Let $\lfloor r \rfloor$ denote the integral part of $r \in \mathbb{R}$ and let $\{r\}$ denote its fractional part. Two valuations $v$ and $v'$ are equivalent[10], denoted $v \sim v'$, iff for all clocks $x$ and $y$:

[10] This is the usual region equivalence [8] with no bound associated to the clocks.

1. $\lfloor v(x) \rfloor = \lfloor v'(x) \rfloor$,
2. $\{v(x)\} = 0$ iff $\{v'(x)\} = 0$,
3. $\{v(x)\} \leq \{v(y)\}$ iff $\{v'(x)\} \leq \{v'(y)\}$.

The following Lemma is an intermediate result for the proof of the Rescheduling Lemma.

**Lemma 3.** *For all non-negative real numbers $t$, $t'$ and $t''$ such that $t > t'$, $t > t''$ and $0 \leq \{t'\} < \{t''\}$ we have:*

$$\{t - t'\} < \{t - t''\} \qquad \text{if} \qquad \{t'\} \leq \{t\} < \{t''\} \tag{2}$$

$$\{t - t''\} < \{t - t'\} \qquad \text{if} \qquad \{t\} < \{t'\} \text{ or } \{t''\} \leq \{t\} \tag{3}$$

*Proof.* First, observe that for non-negative real-numbers $t$ and $t'$:

$$\{t - t'\} = \begin{cases} \{t\} - \{t'\} & \text{if } \{t\} - \{t'\} \geq 0 \\ 1 + \{t\} - \{t'\} & \text{otherwise} \end{cases} \tag{4}$$

Let us first prove (2). From $\{t'\} < \{t''\}$, we have $\{t''\} < \{t'\} + 1$, hence $1 + \{t\} - \{t''\} > \{t\} - \{t'\}$. Then since $\{t'\} \leq \{t\} < \{t''\}$ it comes $\{t - t'\} < \{t - t''\}$ by (4).

Now, we turn to the proof of (3). From $\{t'\} < \{t''\}$ we deduce $\{t\} - \{t'\} > \{t\} - \{t''\}$. If $\{t''\} \leq \{t\}$, from (4) we obtain $\{t - t''\} < \{t - t'\}$. If $\{t\} < \{t'\}$, then further deduce that $1 + \{t\} - \{t'\} > 1 + \{t\} - \{t''\}$ which also lead to $\{t - t''\} < \{t - t'\}$ by (4).

Finally, without loss of generality, we can assume that a run of a timed automaton $\mathcal{B}$ is an alternating sequence of delays $d_i \in \mathbb{R}_{\geq 0}$ and actions $a_i \notin \mathbb{R}_{\geq 0}$: $(\ell_0, v_0) \xrightarrow{d_1} (\ell_0, u_1) \xrightarrow{t_1, a_1} (\ell_1, v_1) \xrightarrow{d_2} (\ell_1, u_2) \xrightarrow{t_2, a_2} \cdots (\ell_n, v_n)$. We omit the timestamps on delays as they are not needed in the sequel.

We are now ready to prove the Rescheduling Lemma. We show that for every open interval $I = (a, b)$ in $(0, 1)$, from every run $\rho = (\ell_0, v_0) \xrightarrow{d_1} (\ell_0, u_1) \xrightarrow{t_1, a_1} (\ell_1, v_1) \xrightarrow{d_2} (\ell_1, u_2) \xrightarrow{t_2, a_2} \cdots (\ell_n, v_n)$ we can build a run $\rho' = (\ell_0, v'_0) \xrightarrow{d_1} (\ell_0, u'_1) \xrightarrow{t'_1, a_1} (\ell_1, v'_1) \xrightarrow{d_2} (\ell_1, u'_2) \xrightarrow{t'_2, a_2} \cdots (\ell_n, v'_n)$ such that $v'_0 = v_0$, and for all $i \in \{1, \ldots, n\}$, if $t_i \in \mathbb{N}$ then $t'_i = t_i$, otherwise, $t'_i \in \lfloor t_i \rfloor + I$, and $v_i \sim v'_i$ and $u_i \sim u'_i$.

*Proof.* We prove by induction on the length of run $\rho$ that all $t'_i$ can be chosen such that $v_i \sim v'_i$ and $u_i \sim u'_i$ for all $i \geq 0$. This is obvious for $v_0$ and $v'_0$ as they are equal. Now, we assume that $v_i \sim v'_i$ holds up to some given $i \geq 0$, and we prove that $u_{i+1} \sim u'_{i+1}$. Observe that this entails $v_{i+1} \sim v'_{i+1}$ as $v_{i+1}$ and $v'_{i+1}$ are obtained from $u_{i+1}$ and $u'_{i+1}$ respectively by resetting the same clocks as specified by transition $a_{i+1}$.

For every clock $x$, let $t^x$ denote the last timestamp before $t_{i+1}$ when clock $x$ has been reset. That is, $t^x$ is the largest timestamp $t_j$ in $\{t_0, \ldots, t_i\}$ such that $x$ is reset on the transition $\sigma_j$. In the same way, we define $t^{x\prime}$ relatively to $t'_{i+1}$. Observe that $u_{i+1}(x) = t_{i+1} - t^x$ and $u'_{i+1}(x) = t'_{i+1} - t^{x\prime}$ for every clock $x$. By induction hypothesis, the lemma holds for $t^x$ and $t^{x\prime}$. That is: if $\{t^x\} = 0$ (i.e. $t^x \in \mathbb{N}$) then $\{t^{x\prime}\} = 0$ otherwise $\{t^{x\prime}\} \in I$. Observe also that $\{u_{i+1}(x)\} = \{u_{i+1}(y)\}$ entails $\{t^x\} = \{t^y\}$ for all clocks $x$ and $y$. The same holds for $u'_{i+1}$, $t^{x\prime}$ and $t^{y\prime}$.

As a first step, we prove that $\lfloor u_{i+1}(x) \rfloor = \lfloor u'_{i+1}(x) \rfloor$ for every clock $x$ which corresponds to condition 1 of the region equivalence. We prove that this holds for any choice of $t'_{i+1}$ that respects the conditions in the lemma. We have $u_{i+1}(x) = \lfloor t_{i+1} \rfloor - \lfloor t^x \rfloor + \{t_{i+1}\} - \{t^x\}$ and $u'_{i+1}(x) = \lfloor t'_{i+1} \rfloor - \lfloor t^{x\prime} \rfloor + \{t'_{i+1}\} - \{t^{x\prime}\}$. The cases where $\{t^x\} = 0$ or $\{t_{i+1}\} = 0$ are straightforward. We only detail the case where $\{t^x\} \in (0;1)$, which entails $\{t^{x\prime}\} \in I$ by induction, and $t_{i+1} \in (0;1)$. We show that any choice of $\{t'_{i+1}\} \in I$ is valid. We have: $\lfloor t_{i+1} \rfloor - \lfloor t^x \rfloor - 1 < \lfloor u_{i+1}(x) \rfloor < \lfloor t_{i+1} \rfloor - \lfloor t^x \rfloor + 1$ and $\lfloor t'_{i+1} \rfloor - \lfloor t^{x\prime} \rfloor + a - b < \lfloor u'_{i+1}(x) \rfloor < \lfloor t'_{i+1} \rfloor - \lfloor t^{x\prime} \rfloor + b - a$ (recall $I = (a, b)$). Now, since $\lfloor t_{i+1} \rfloor = \lfloor t'_{i+1} \rfloor$, $\lfloor t^x \rfloor = \lfloor t^{x\prime} \rfloor$, and 0 is the only integer between $a - b$ and $b - a$, it comes $\lfloor u_{i+1}(x) \rfloor = \lfloor u'_{i+1}(x) \rfloor$.

In a second step, we prove that conditions 2 and 3 of the region equivalence hold. Let $X_0, \ldots, X_k \subseteq X$ define a partition of the clocks according to their fractional part in the valuation $v_i$. Formally, for each $x, y \in X_j$, $\{v_i(x)\} = \{v_i(y)\}$, for each $x \in X_j$ and $y \in X_{j-1}$, $\{v_i(y)\} < \{v_i(x)\}$, and $\bigcup_{j=0}^{k} X_j = X$. Observe that $v_i$ and $v'_i$ define the same partition of clocks as $v_i \sim v'_i$. This partition is depicted in Figure 5 to the left. As time elapses from $v_i$ and $v'_i$, the fractional part of clock valuations increases and the ordering of partitions changes. Some clocks, say $X_0, \ldots, X_{j-1}$ have their fractional part increased, whereas some others, say $X_j \ldots, X_k$ have their fractional part decreased as they have been set back to 0 meanwhile. Assume that the ordering of fractional part of the clocks in $u_{i+1}$ is as depicted in Figure 5 to the right. We now show that $\{t'_{i+1}\}$ can always be chosen in such a way that $u'_{i+1}$ has the same ordering of the fractional part of the clocks as $u_{i+1}$, which will conclude the proof that $u_{i+1} \sim u'_{i+1}$.

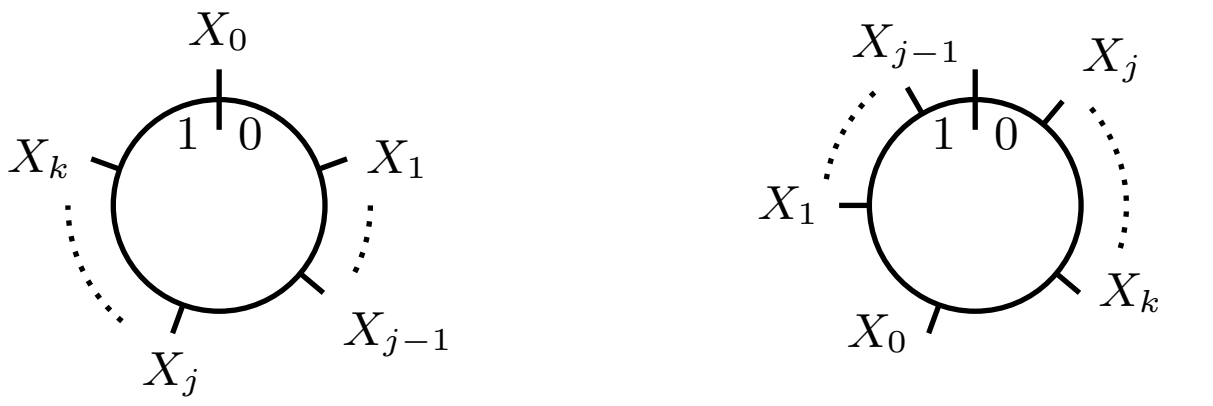

Fig. 5: The ring of fractional parts before (left) and after (right) time elapses.

We first consider the case when the partition only contains the single set $X$. As all the clocks have the same fractional part, only condition 2 of the region equivalence needs to be considered. If $\{u_{i+1}(x)\} = 0$ for all clock $x$, we choose $\{t'_{i+1}\} = \{t^{x'}\}$ which yields $\{u'_{i+1}(x)\} = 0$. By induction, $\{t'_{i+1}\}$ satisfies the lemma. Conversely, when $\{u_{i+1}(x)\} > 0$ for all clock $x$, choosing $\{t'_{i+1}\} \neq \{t^{x'}\}$ guarantees that $\{u'_{i+1}(x)\} > 0$ too. We need to show that there always exists such a solution. From $\{u_{i+1}(x)\} > 0$, we obtain $\{t_{i+1} - t^x\} > 0$, hence we cannot have $\{t_{i+1}\} = 0$ and $\{t^x\} = 0$ at the same time. If $\{t_{i+1}\} = 0$ then $\{t^x\} > 0$, hence $\{t'_{i+1}\} = 0$ is a solution since $\{t^{x'}\} > 0$ by induction hypothesis. Conversely, if $\{t_{i+1}\} \in (0;1)$, then we can choose any $\{t'_{i+1}\} \in I$ distinct from $\{t^{x'}\}$ (recall $\{t^{x'}\} = \{t^{y'}\}$ for all clocks $x$ and $y$).

Now, we consider a partition $X_0, \ldots, X_k$ of the clocks in $v_i$ and $v'_i$, with $k \geq 1$, and the partition $X_j, \ldots, X_k, X_0, \ldots, X_{j-1}$ in $u_{i+1}$ as depicted in Figure 5. Let us first focus on the case when $\{u_{i+1}(x)\} = 0$ for $x \in X_j$. As $u'_{i+1}(x) = t'_{i+1} - t^{x'}$, for $\{u'_{i+1}(x)\} = 0$ it must be the case that $\{t'_{i+1}\} = \{t^{x'}\}$. By induction hypothesis, this value of $\{t'_{i+1}\}$ satisfies the lemma.

Now consider the case where $\{u_{i+1}(x)\} > 0$ for $x \in X_j$. As seen on Figure 5 to the right, we need to make sure that, in valuation $u'_{i+1}$, the clocks in $X_j$ have the smallest fractional part and the clocks in $X_{j-1}$ have the biggest one. This is ensured by condition $\{u'_{i+1}(x)\} < \{u'_{i+1}(y)\}$ for $x \in X_j$ and $y \in X_{j-1}$, which translate as:

$$\{t'_{i+1} - t^{x'}\} > 0 \quad \text{and} \quad \{t'_{i+1} - t^{x'}\} < \{t'_{i+1} - t^{y'}\} \tag{5}$$

We distinguish two cases whether $\{t^{y'}\} > \{t^{x'}\}$ or $\{t^{y'}\} < \{t^{x'}\}$. Let us consider the first case. From Lemma 3 and (5), we need to find a value of $\{t'_{i+1}\}$ such that $\{t^{x'}\} < \{t'_{i+1}\} < \{t^{y'}\}$. By induction hypothesis we have $\{t^{y'}\} \in I$ and the following two cases for $\{t^{x'}\}$:

- either $\{t^{x'}\} = 0$, then $\{t^x\} = 0$ by induction, hence $\{t_{i+1}\} > 0$ as $\{u_{i+1}(x)\} > 0$. Since $\{t_{i+1}\} \in (0;1)$ we must choose $\{t'_{i+1}\}$ in $I$. Taking $\{t'_{i+1}\} = \frac{a+\{t^{y'}\}}{2}$ fulfills all the requirements.
- or $\{t^{x'}\} \in I$. Then choosing $\{t'_{i+1}\} = \frac{\{t^{x'}\}+\{t^{y'}\}}{2}$ yields a solution.

It remains to consider the case when $\{t^{y'}\} < \{t^{x'}\}$. Applying Lemma 3 on (5) yields two sets of solutions: $\{t'_{i+1}\} < \{t^{y'}\}$ or $\{t^{x'}\} \leq \{t'_{i+1}\}$.

- If $\{t_{i+1}\} \in (0;1)$, then $\{t'_{i+1}\} = \frac{\{t^{x'}\}+b}{2}$ is a solution as $\{t^{x'}\} \leq \{t'_{i+1}\}$ and, by induction hypothesis, $\{t^{x'}\} \in I$ since $\{t^{y'}\} < \{t^{x'}\}$ (i.e. $\{t^{x'}\} \neq 0$).
- Now, if $\{t_{i+1}\} = 0$ we have $\{t^y\} > 0$. Indeed, as $y \in X_{j-1}$, we have $\{u_{i+1}(y)\} = \{t_{i+1} - t^y\} > 0$ and $\{t^y\} = 0$ entails $\{t_{i+1}\} > 0$, a contradiction. By induction hypothesis, from $\{t^y\} > 0$ we get $\{t^{y'}\} > 0$. Hence, we can pick $\{t'_{i+1}\} = 0$ which satisfies $\{t'_{i+1}\} < \{t^{y'}\}$.

Finally, it remains the case when the ordering of fractional parts is the same in $v_i$ and $u_{i+1}$. Then, considering $X_j = X_0$, and $X_{j-1} = X_k$ yields a solution for $\{t'_{i+1}\}$ as stated above.

### D.2 Abstraction of communicating timed automata with emptiness tests is difficult

In this section we discuss why our abstraction (presented in Section 4) does not work with emptiness tests and why it seems difficult to find a suitable abstraction that preserves the topology. Notice that an abstraction that does not preserve the topology is known for the particular case of a channel with distinct sender and receiver [18]

*Our construction is not sound for emptiness test* We propose the simple example in Figure 6. From top to bottom, there are a sender and a receiver, communicating via a channel $c$. We can easily verify that there is no global run in this system. Indeed, the actions along a global run have to be in the following order: $c!a; c?a; c == \varepsilon; c!a$. Then the emptiness test cannot be satisfied as $c$ is not empty. Hence the receiver cannot reach its final location. On the contrary, the system of communicating tick automata obtained by applying the construction in Section 4 has a global run that reaches the final locations. This system is depicted in Figure 7. The global run corresponds to the sequence of actions $c!a; c?a; c == \varepsilon; \tau; c == \varepsilon; c!a; c?a$ where both processes synchronize on $\tau$. Observe that this global run cannot be re-scheduled in the spirit of the Rescheduling Lemma. Indeed, both real-time constraints and dependencies between the communication actions prevent to swap actions $c == \varepsilon; c!a$ as $c!a; c == \varepsilon$.

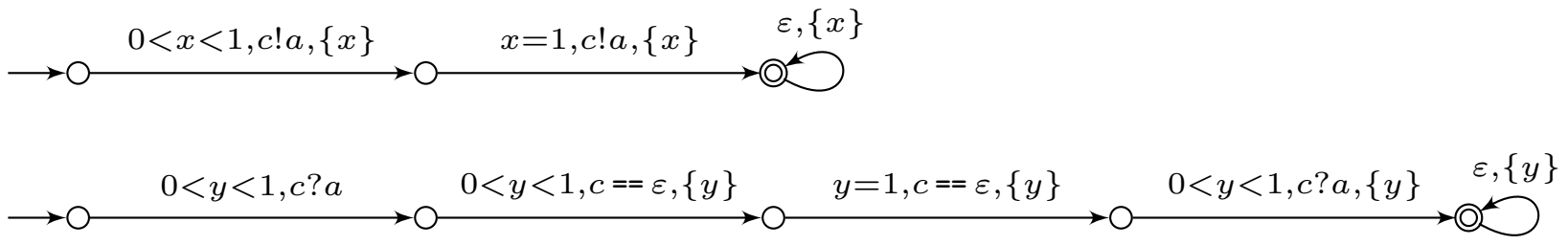

Fig. 6: A counter-example to our abstraction with emptiness test.

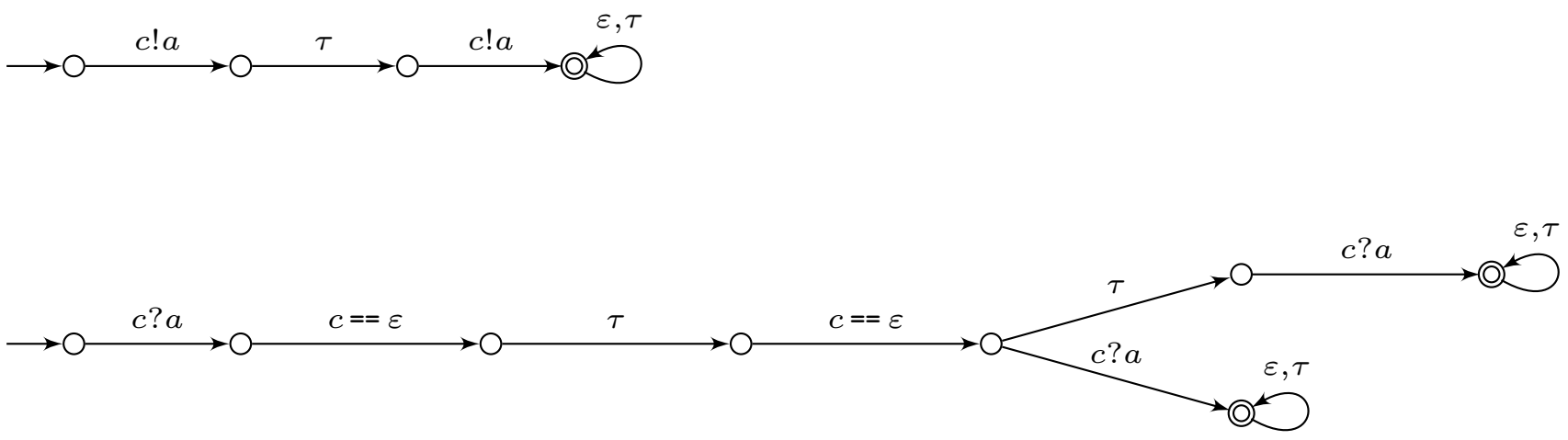

Fig. 7: A counter-example to our abstraction with emptiness test.

*Why soundness is hard to achieve* Our abstraction is based on the possibility to define a partition scheduling that allocates one slot per time unit (the interval $I$ in the Rescheduling Lemma) to each process in the system. In the previous section, we have seen that in presence of emptiness test, one slot per process may not be sufficient. We now show that we cannot even find a bound on the number of slots per time unit needed by each process.

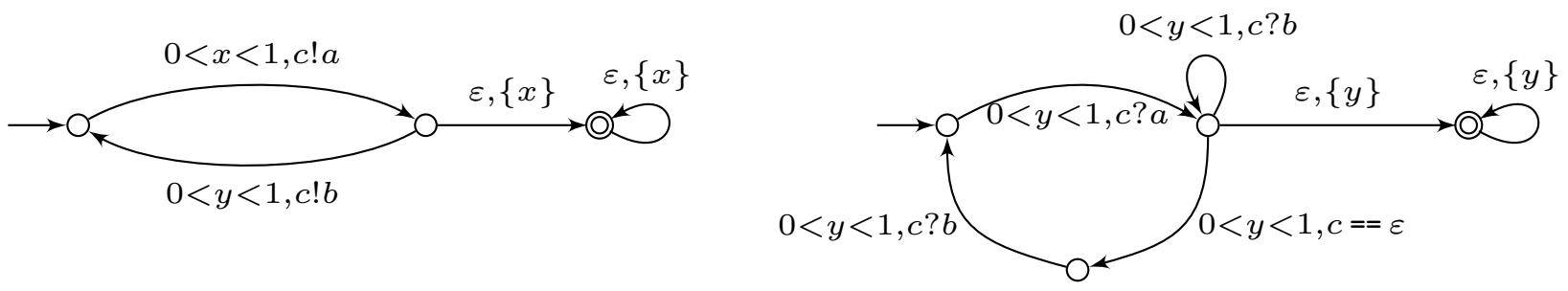

Fig. 8: A counter-example to our abstraction with emptiness test.

Figure 8 shows an example with a sender $p$ (left) and a receiver $q$ (right) that communicate via a channel $c = (p, q)$. Consider a global run of the system where the sender $p$ performs actions $c!a; c!b$ while the receiver $q$ does actions $c?a; c == \varepsilon; c?b$. Obviously, $q$ has to perform the emptiness test $c == \varepsilon$ between the two emissions by $p$. Observe that both processes can iterate this behavior. Finally, all these actions occur in one time unit. This shows that the number of slots needed by $p$ and $q$ depends on the number of iterations on their respective loops. Thus there may not be an uniform choice of slots in presence of emptiness tests.

Notice that this is due to a convergence phenomenon but not necessary to Zeno behaviors. Adding loops that reset the clocks on the initial locations of both process, we could let one time unit elapse infinitely often, but the problem would remain the same.